\documentclass[%
nofootinbib,
 amsmath,amssymb,
 aps,
 prfluids,
]{revtex4-2}

\usepackage{graphicx}
\usepackage{dcolumn}
\usepackage{bm}
\usepackage{hyperref}

\usepackage{booktabs}    
\usepackage{subcaption}  
\usepackage{xcolor}      
\usepackage[normalem]{ulem}

\begin{document}


\title{Physics-based distinction of nonequilibrium effects in near-wall modeling of turbulent separation bubble with and without sweep}

\author{Imran Hayat}
\author{George Ilhwan Park}%
 \email{Corresponding author: gipark@seas.upenn.edu}
\affiliation{
 University of Pennsylvania, Philadelphia, Pennsylvania 19104
}

\date{\today}

\begin{abstract}

Pressure-gradient-induced separation of swept and unswept turbulent boundary layers, based on the DNS studies of Coleman et al. (J. Fluid Mech. (2018), vol. 847, 28-70, and J. Fluid Mech. (2019), vol. 880, 684–706), have been analyzed for various nonequilibrium effects. The goal is to isolate physical processes critical to near-wall flow modeling. The decomposition of skin friction into contributing physical terms, proposed by Renard and Deck (J. Fluid Mech. (2016), vol. 790, 339-367) (short: RD decomposition), affords several key insights into the near-wall physics of these flows. In the unswept case, spatial growth term (encapsulating nonequilibrium effects) and TKE production appear to be the dominant contributing terms in the RD decomposition in nonequilibrium zones (i.e., separated zone and upstream/downstream pressure-gradient zones). Inspection of wall-normal profiles of these terms reveals that only the spatial growth term dominates the skin-friction balance in the inner layer close to the separation bubble, implying a strong need for incorporating nonequilibrium terms in wall-flux modeling of the unswept separation bubble. The comparison of streamwise RD decomposition of swept and unswept cases shows that a larger accumulated Clauser-pressure-gradient parameter history in the latter enhances the outer-layer TKE production peak in the adverse pressure gradient zone. This energization of outer (large-scale) dynamics appears to be the reason for the diminished separation bubble size in the unswept case compared to the swept case. The spanwise RD decomposition in the swept case indicates that the downstream spanwise boundary layer, including that within the separation bubble, largely retains the upstream zero-pressure-gradient characteristics. This seems to ease the near-wall modeling challenge in the separated region, especially for basic models with an inherent log-law assumption. Wall-modeled LES of the swept and unswept cases are then performed using three wall models with varying physical fidelity. Many of the above modeling implications from the DNS are verified by analyzing various flow quantities. In particular, the extension of RD decomposition to wall models underpins the criticality of spatial growth term in the vicinity of the separation bubble, and the corresponding superior predictions by the PDE wall model
due to its accurate capturing of this term.

\end{abstract}

\maketitle

\section{\label{sec:intro}Introduction}
Most turbulent boundary layers (TBL) encountered in
nature and engineering flows exhibit nonequilibrium effects such as pressure gradient, mean three-dimensionality, and flow separation, among others. Examples include atmospheric TBL over natural obstacles and flow over swept-wing aircraft, turbine blades, and marine vehicle hulls. Despite the ubiquity of nonequilibrium effects, most theories and models of wall turbulence developed so far are based on the zero-pressure-gradient (ZPG) or near-ZPG equilibrium TBL with mean profiles that are mainly one-dimensional. In the present study, we investigate the Direct Numerical Simulations (DNS) dataset of idealized (infinite) unswept and swept wings, where 2D and 3D TBLs are subjected to both favorable pressure gradient (FPG) and adverse pressure gradient (APG), with the APG inducing flow separation at the wall. All these nonequilibrium effects severely challenge many of the modeling assumptions inherent in the equilibrium 2D-TBL models. Additionally, the prediction of TBL separation, especially through scale-resolving simulations such as wall-modeled LES (WMLES), is of prime importance to the aerospace community \cite{slotnick2014cfd} due to the critical role of separation in setting the optimal performance conditions of an aircraft. Foregoing in mind, we also conduct WMLES of the reference DNS configurations using three different wall models, with the primary objective of validating the physical insights and modeling implications derived from analyzing the DNS datasets.

Various studies in the past have investigated nonequilibrium TBLs in complex flows using WMLES. \citet{rezaeiravesh2019} systematically studied turbulent channel flow using an algebraic wall model, to investigate the predictive accuracy of WMLES based on various wall-model parameters such as grid resolution, matching height, SGS model, and log-law parameters. Similarly, \citet{wang2020} performed a comparative study of various wall models and SGS models in high Reynolds number turbulent channel flow. \citet{vane2013} and \citet{iyer2017} performed WMLES for nonequilibrium shock-TBL interactions in a normal shock and an oblique shock flow, the latter of which induced separation. \citet{iyer2016,iyer21,iyer2023} carried out a campaign of WMLES to study smooth-body separation over 2D and 3D bumps using equilibrium and nonequilibrium wall models. \citet{KawaiAsada2013} studied transitional and separated flow over an airfoil near stall condition using a nonequilibrium wall model, while \citet{lehmkuhl2018} and \citet{goc2021} conducted WMLES for full aircraft geometry using algebraic wall models. The vast majority of studies in the WMLES literature (including the aforementioned investigations) simply present a posteriori evaluation of WMLES results against DNS or experiments. Although useful to develop databases and practitioner guidelines for WMLES, such studies are limited in their scope to provide a fundamental understanding of the underlying mechanisms of turbulence models. In particular, it is imperative to develop and investigate turbulence models via a physics-based analysis framework, as emphasized by the NASA 2030 CFD Vision Report \cite{slotnick2014cfd}. Such physics-based approaches can highlight the shortcomings of models in different flows more fundamentally and inform necessary modifications to the models more systematically. These physics-based analysis frameworks are especially crucial in nonequilibrium TBLs such as the present flow, where various distinct nonequilibrium effects are superimposed and possibly interact with one another.

A few studies in the past couple of decades have attempted to compare wall models in nonequilibrium TBLs, with an emphasis on the internal physical mechanisms of the models. \citet{wang02} systematically incorporated various nonequilibrium (pressure gradient, advection, and unsteady) terms in the TBL-equation-based wall models and tried to quantify the effect of each term on the predicted skin friction of unsteady trailing-edge separation. Importantly, they proposed a dynamic correction to the model based on a physics-based analysis, which revealed a partial resolution of the Reynolds stresses by the wall model when advective terms were incorporated. \citet{park17aiaa} compared the ODE equilibrium wall model (EQWM) and the PDE non-equilibrium wall model (PDE NEQWM) in a separating flow over a wall-mounted hump and showed, via the analysis of total shear stress profiles and the momentum budget within the wall model, that the PDE NEQWM outperforms the EQWM in the separation bubble and the recovery region. \citet{hu23} performed a similar budget analysis of three wall models (EQWM, PDE NEQWM, and integral NEQWM) in a 30$^\text{o}$ bent duct. They showed separable contributions to the wall shear stress direction from the equilibrium stress part and the integrated non-equilibrium effects, with the latter found to differ among wall models. Similarly, \citet{tamaki2020} employed the Von-Karman integral (VKI) momentum equation framework to examine the contributions from skin friction, Reynolds stresses, and acceleration/deceleration terms to the momentum-thickness development in an airfoil flow with trailing-edge separation. They argued that contributions from skin friction were only important in the far upstream APG region and became insignificant close to the separation.

It is well known that, in addition to the direct influence of near-wall turbulence on the mean wall shear stress, the large-scale dynamics in the outer layer play a key role in modulating the near-wall dynamics \cite{hutchins2007,mathis2009} and therefore, generating the  mean wall shear stress. Based on the foregoing view, several integral frameworks have been developed for TBLs to decompose the mean skin friction coefficient (or the wall-shear stress) into contributions from physical processes across the entire boundary layer thickness. The main idea is that these physical processes represent mechanisms for skin friction generation and therefore a causal relation exists between them. One of the first among these was the so-called FIK identity developed by \citet{fukagata2002}. Based on the triple integration of the averaged streamwise momentum equation, this analysis identified four contributing components to the mean skin friction (laminar, turbulent, inhomogeneous, and transient components). Although the FIK identity has been applied successfully to several internal flows, it is not necessarily effective in isolating the contributing terms into the wall-stress-generating physical phenomena for spatially-developing boundary layers \cite{RD2014,elnahhas2022}. Furthermore, \citet{RD2016} argued that it lacks a causal relation for skin friction due to its physically uninterpretable aspects, such as the linearly-weighted Reynolds stresses and the three successive integrations.
\citet{yoon2016} introduced a vorticity-transport-based decomposition of skin friction and explained the influence of vortical structures on skin friction through the contributing terms. However, this correlation was also derived from a triply integrated mean spanwise vorticity equation, making it cumbersome to use as a diagnostic tool. Recently, \citet{elnahhas2022} devised an angular-momentum integral (AMI) based decomposition technique. 
This method can be sensitive to the definition of $\delta_{99}$, and although it was mitigated in the original AMI study by employing the freestream definition of \citet{griffin2021}, the sensitivity persisted in the present flow. To overcome the aforementioned shortcomings of previous frameworks and provide a simple and effective diagnostic tool for understanding the flow physics of nonequilibrium flows, in the present study we employ an alternative skin-friction decomposition based on the energy budget, called the RD decomposition (due to \citet{RD2016}).

The RD decomposition offers several advantages over the frameworks described above. First, by adopting the energy budget instead of the momentum budget (used in VKI, FIK, and AMI frameworks), a single integration of the balance equation is made possible without eliminating the relevant contributing terms. In contrast, a single wall-normal integration of the momentum equation across the boundary layer in other frameworks causes the Reynolds shear stress term to vanish (as shown in Eq.(1.3) and (1.4) of \cite{RD2016}). 
Consequently, the FIK analysis must resort to the non-physical triple integration \cite{RD2016} that results in a linearly-weighted integral of the Reynolds shear stress. Although the linear weight somewhat fits the physical description of the near-wall flow (in terms of the increasing influence of the Reynolds-shear-stress-carrying fluctuations on the mean skin friction close to the wall), it does not represent a physically valid profile \cite{RD2016}. 
The corresponding term in the RD decomposition is the integral of the turbulent kinetic energy production or equivalently, the Reynolds shear stress weighted by the mean velocity gradient, which increases with the decreasing wall distance. Therefore, the RD decomposition provides a physically interpretable contribution of turbulent fluctuations to the mean skin friction. Lastly, unlike the FIK identity where the upper limit of integrals is restricted to 
the boundary-layer height ($y=\delta_{99}$), the RD decomposition can account for the Reynolds shear stress outside the nominal boundary layer thickness \cite{RD2016}. Furthermore, the RD decomposition is found to be extremely robust to the definition of $\delta_{99}$ in contrast to the AMI-based approach where the decomposed terms showed a strong sensitivity to the definition of $\delta_{99}$, at least for the present flow. Additionally, the RD decomposition can be straightforwardly adapted to the near-wall region only, in the context of wall modeling. This is afforded by the single integration required in this technique, in contrast to other techniques with multiple integrations that give rise to complicated (and often physically uninterpretable) terms. 

Given the physics-based interpretability of its constituent terms, the RD decomposition is particularly useful when analyzing the mechanisms of success or failure of wall-stress models for LES, as it provides a way to identify and isolate the most critical terms contributing to skin friction. This enables a physics-based diagnostic analysis of the wall-model performance, by noting how well the model captures the most significant contributing terms in the RD decomposition. In the present study, this approach lends itself to distinguishing wall models based on their characterization of the nonequilibrium contributions in the near-wall region. We note that the RD decomposition has been applied to various flow scenarios in the past, to investigate physical phenomena in the flow and their contribution to skin friction. These applications have included compressible channel flow \cite{fan2019a,li2019} and turbulent boundary layer in ZPG \cite{fan2019b} and APG \cite{fan2020}, for example. Two aspects of the present work distinguish it from prior studies; first, the RD decomposition is extended for the first time to TBLs with mean three-dimensionality and separation; second, the use of RD decomposition to analyze differences in the physics of different wall models is unprecedented, to the best of our knowledge.

The objectives of this paper are: 1) conducting a physics-based analysis of nonequilibrium effects via the RD decomposition applied to the DNS datasets, and 2) adapting the RD decomposition to the near-wall region in the WMLES and applying insights from the RD decomposition in the DNS, to improve our understanding of the mechanisms of success/failure in wall models. The remainder of the paper is organized as follows. The flow configurations are described in section \ref{sec:flow_config}. Analyses of the flow physics based on the DNS datasets are detailed in section~\ref{sec:DNS_analysis}. The numerical setup of WMLES and its results and analysis are provided in section \ref{sec:WMLES}. Conclusions are offered in section \ref{sec:conclusion}.







\section{Flow configuration}\label{sec:flow_config}

\begin{table}[t]
\caption{\label{tab:table1} Flow conditions and computational domain information for the reference DNS and the present WMLES. $\sigma$ is the nominal flow angle with $x$ at the inlet; $U_{\infty}\theta_{x_{0}}/\nu$ is the momentum thickness-based Reynolds number at $x=0$; $U_{\infty}Y/\nu$ is the nominal Reynolds number based on the domain height; $L_{x}/Y$, $L_{y}/Y$, and $L_{z}/Y$ are the domain extents in $x,y$, and $z$ directions, respectively; $x_{\rm in}$ and $x_{\rm out}$ are the inlet and outlet locations (relative approximately to the center of the separation bubble); $x_{\rm sponge}$ is the starting location of the sponge region shown in Fig.~\ref{fig:schematic_sepbub}($a$).
}
\centering
\begin{ruledtabular}
\begin{tabular}{lccccccccc}
Case & $\sigma$ (deg.) & $U_{\infty}\theta_{x_{0}}/\nu $ & $U_{\infty}Y/\nu $ & $L_{x}/Y$  & $L_{y}/Y$ & $L_{z}/Y$ & $x_{\rm in}$ & $x_{\rm out}$ & $x_{\rm sponge}$\\\hline
C0 (DNS)  & 0  & 3057 & 80000 & 26.0 & --  & 4.0 & -15.35 & 10.65 & -- \\
C0 (WMLES)& "  &   "  &   "   & 33.5 & 1.0 &   " & -17.85 & 15.65 & 10.0\\
C35 (DNS)  & 35 & 3031 & 80000 & 26.0 & --  & 4.0 & -13.63 & 12.37 & --  \\
C35 (WMLES)& "  &   "  &   "   & 30.5 & 1.0 &   " & -14.85 & 15.65 & 10.0\\
\end{tabular}
\end{ruledtabular}
\end{table}

\begin{figure}[t]
\centering
\begin{subfigure}[b]{\textwidth}
    \includegraphics[trim=0 130 0 150,clip,width=\textwidth]{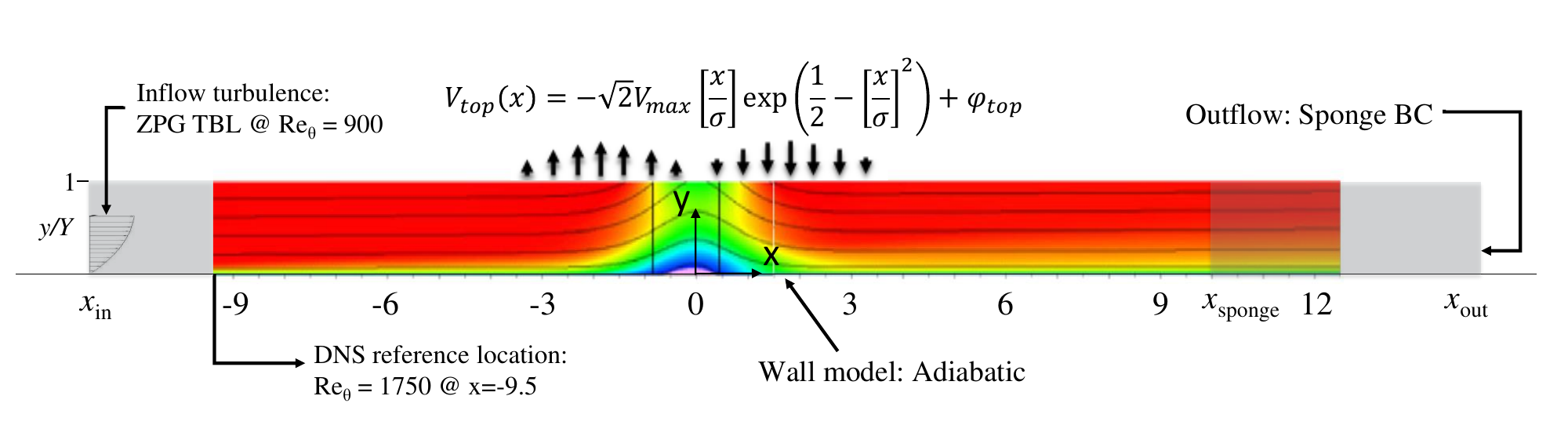}
    \caption{}
\end{subfigure}
~
\begin{subfigure}[b]{0.6\textwidth}
    \includegraphics[trim=0 0 0 350,clip,width=\textwidth]{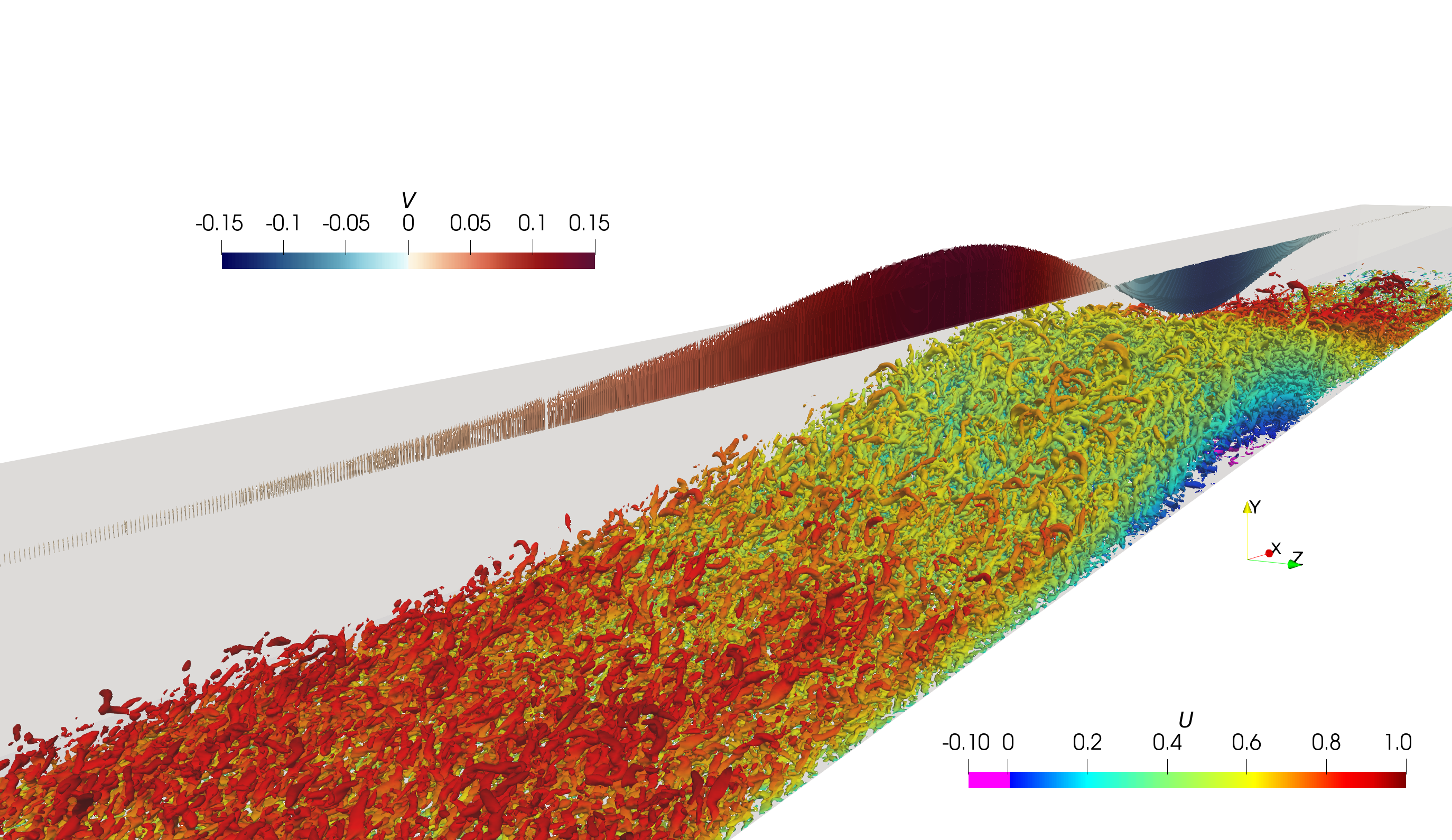}
    \caption{}
\end{subfigure}
\caption{($a$) 
Schematic of the computational domain used and the boundary conditions imposed in the present WMLES of separated TBL. Mean velocity contours from the DNS of \citet{Coleman2018} are shown for reference, where the magenta color shows the reversed mean flow region. Shaded regions depict the inflow development zone and the outflow sponge region in WMLES. ZPG reference location is shown for Case C0. For Case C35, the reference location is at $x_{\rm ref}/Y=-8.5$. ($b$) Iso-contours of Q criterion colored by mean streamwise velocity, in the WMLES of Case C0. The transpiration profile at the top boundary is depicted through vertical velocity vectors along the centerline.}
\label{fig:schematic_sepbub}
\end{figure}

\begin{figure}[t]
\centering
\includegraphics[trim=30 170 10 100,clip,width=\textwidth]{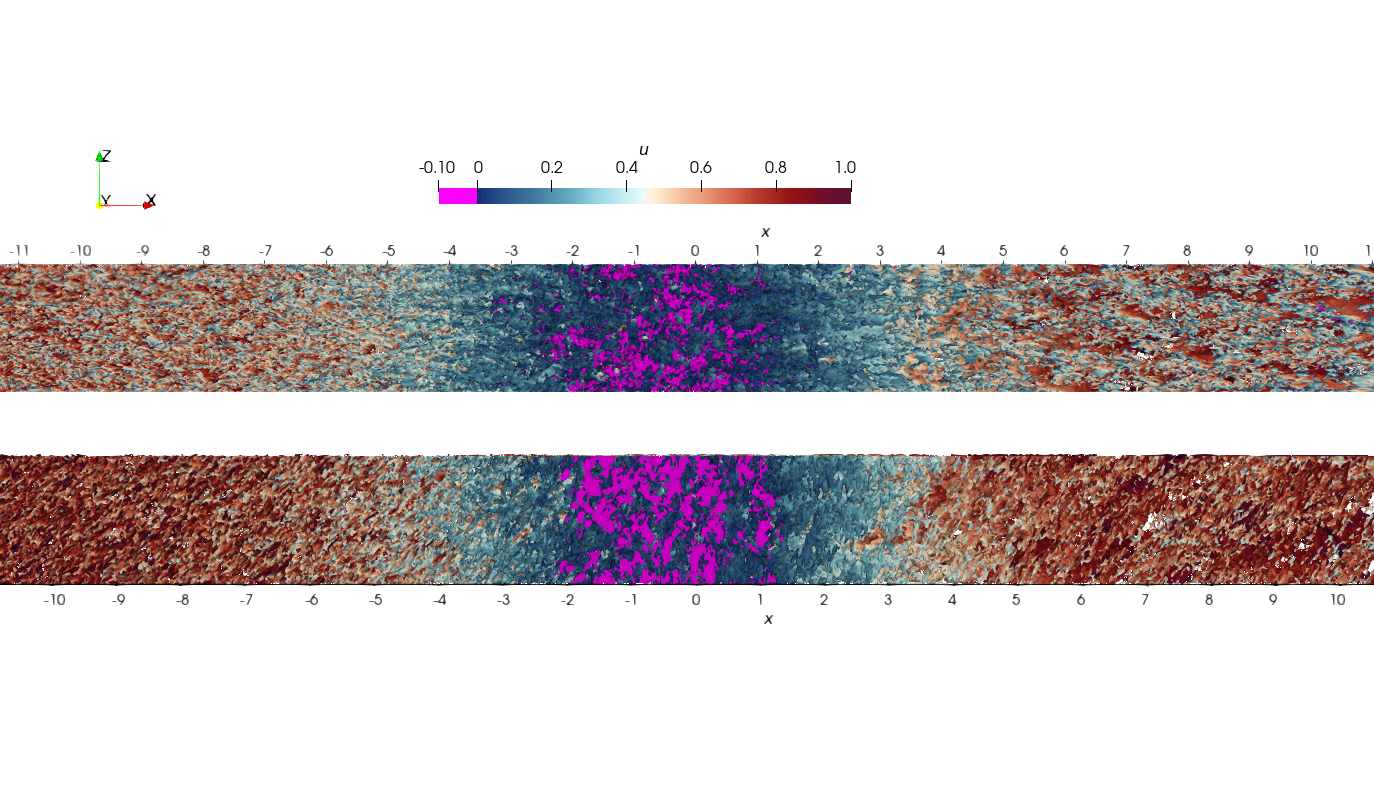}
\caption{Near-wall flow comparison between C0 (top) and C35 (bottom) cases from WMLES (using PDE nonequilibrium wall model). Isocontours of spanwise velocity fluctuations ($w'/U_{\infty}$=0.005) colored by mean streamwise velocity. In both cases, the pressure gradient vector is aligned with the horizontal direction ($x$). 
The inflow freestream is at angles of 0 and 35 degrees in C0 and C35, respectively.
} \label{fig:flow_C0_vs_C35}
\end{figure}

The flow configuration in the present study is based on the DNS of turbulent separation bubble by \citet{Coleman2018,Coleman2019}. The reference studies feature two variants of the flow, one without a spanwise sweep (henceforth denoted as Case C0), and the other with a spanwise sweep (denoted as Case C35). The sweep in Case C35 is introduced through a spanwise velocity component imposed at the inflow plane. As a result, the incoming freestream flow is inclined at an angle of $35^{\circ}$ to the chordwise ($x$) axis in the $x-z$ plane, while a chordwise pressure gradient identical to the unswept case is maintained. Therefore, a 3D-TBL is expected in Case C35 due to the non-alignment of the pressure gradient vector and the freestream. These two cases allow us to systematically investigate various nonequilibrium effects in a simple 2D-TBL separation in Case C0, and in a more realistic 3D-TBL separation that is characteristic of flow over swept wings, in Case C35. 
In both configurations, an incompressible ZPG turbulent boundary layer on a flat plate is subjected to an APG followed by an FPG. As a result, the flow separates and then quickly reattaches a small distance downstream. The APG and FPG are induced through suction and blowing imposed on a wall-parallel plane at a wall-normal distance of $y=Y$ from the bottom wall, with the direction of transpiration changing at $x=0$, as shown in Fig.~\ref{fig:schematic_sepbub}. The transpiration profile imposed at this plane is given by,
 
\begin{equation}
\label{eqn:transpiration}
    V_{top}(x) = -\sqrt{2} V_{max} \left[\frac{x}{\epsilon}\right] \text{exp} \left(\frac{1}{2} - \left[\frac{x}{\epsilon}\right]^{2}\right)  + \varphi_{top} ,
\end{equation}

\noindent where $V_{max} = 0.1333$, $\epsilon = 3.66$, and $\varphi_{top} = 0.0034$ for the case under consideration in the present study (Case C in \cite{Coleman2018}). Table \ref{tab:table1} provides information on the nominal flow conditions at reference locations and the domain extent in the DNS and WMLES. 
Figure~\ref{fig:flow_C0_vs_C35} provides an overview of the flow and structural differences between C0 and C35 cases, by visualizing the near-wall flow. Regions of flow acceleration, separation, and deceleration are clearly visible in this figure. Also noticeable is the general orientation of flow structures, which is along the direction of the pressure-gradient vector in Case C0, and swept at an angle that keeps increasing and becomes nearly perpendicular to the pressure-gradient vector within the separation bubble, in Case C35. Further details on the flow characteristics of these two cases can be found in \cite{Coleman2018} and \cite{Coleman2019}. Note that the same coordinate system used in the reference study is adopted in the present study, with $(x,y,z)$ representing the streamwise (or chordwise in Case C35), wall-normal, and spanwise directions, respectively. The instantaneous and mean velocity components in these directions are denoted by $(u,v,w)$ and $(U,V,W)$, respectively.

 A few remarks are in order regarding the suitability of the reference studies \cite{Coleman2018} and \cite{Coleman2019} to analyze nonequilibrium TBLs. Multiple nonequilibrium effects are at play in this flow, namely, pressure gradient, flow acceleration/deceleration, flow separation, and mean three-dimensionality. 
 Furthermore, the strength of each of these effects varies in different regions of the flow, both in the streamwise and wall-normal directions. Therefore, the two reference cases provide reasonably complex and realistic conditions to develop and test an analysis framework that can effectively isolate nonequilibrium effects in the flow. Later, this will be useful when analyzing the performance of wall models in predicting nonequilibrium effects in the present flow. 

\section{Analysis of the DNS dataset for nonequilibrium effects}\label{sec:DNS_analysis}

In this section, we attempt to dissect the flow physics of C0 and C35 cases by applying various skin-friction decomposition techniques based on the integral boundary layer relation, to the DNS data. The goal is to first validate these analysis frameworks in the present flow through the DNS and then show their effectiveness in isolating nonequilibrium effects. Later, we will extend the more effective analysis framework to the WMLES results in Section~\ref{sec:WMLES}, to understand the underlying mechanisms of wall models in predicting nonequilibrium effects.\\

\subsection{Integral analysis of flow physics and decomposition of mean skin friction}\label{sec:flow_phy}

We begin with the most basic integral relation, the Von-Karman integral (VKI) boundary layer equation \cite{karman21} given by,

\begin{equation}\label{eq:VKI}
\frac{d\theta}{dx} = I + II + III \;,\; I \equiv \frac{C^{*}_{f}}{2} \;,\; II \equiv -(2+H)\frac{\theta}{U_{e}}\frac{d U_{e}}{dx} \;,\; III \equiv  \frac{1}{U^2_{e}} \frac{d}{dx}  \int_{0}^{Y} \langle u'u' \rangle dy,
\end{equation}

\noindent where $\theta$ is the momentum thickness, $U_e$ is the mean streamwise velocity at the local boundary-layer edge, $C^{*}_{f} = 2\tau_w/(\rho_\infty U^2_e)$ is the mean skin friction based on $U_e$ ($\tau_w$ being the mean wall shear stress), $H=\delta^*/\theta$ is the shape factor ($\delta^*$ being the displacement thickness), and $\langle u'u' \rangle$ is the streamwise Reynolds normal stress component. Throughout this paper, $\langle \cdot \rangle$ denotes averaging of a quantity in time and spanwise direction, with $'$ denoting fluctuations about the mean. In particular for velocity, $\langle u \rangle$ and $U$ are used interchangeably to denote mean streamwise velocity, and so on for other components. Note that all thicknesses and the edge velocity $U_e$ in the present study are based on the generalized vorticity-based velocity of \citet{Coleman2018}. The integral equation (\ref{eq:VKI}) was also employed by \citet{tamaki2020} to analyze the  contribution of skin friction to boundary layer development in a near-stall airfoil flow, 
although only terms $I$ and $II$ were used therein . 
Note that term $I$ represents contributions from the local skin friction whereas term $II$ encapsulates the contributions from acceleration and deceleration in the freestream. In the present study, term $III$ is added to the right-hand side (RHS) to account for the streamwise heterogeneity of the Reynolds normal stress, which is non-negligible in the present flow, especially in the vicinity of the separated region. In fact, \citet{Coleman2018} showed (in Fig. 17($b$) of that paper) that the inclusion of term $III$ was necessary to close the momentum budget when the boundary-layer approximation was applied to Case C0. It is worth noting that Eq. (\ref{eq:VKI}) can be derived by differentiating the streamwise and wall-normal-integrated momentum balance used in Fig. 17($b$) of \cite{Coleman2018}. The steps for this derivation are provided in Appendix \ref{appendix:VKI_derivation}. 

\begin{figure}[t]
\centering
\begin{subfigure}[b]{0.49\textwidth}
    \includegraphics[trim=100 400 780 300,clip,width=\textwidth]{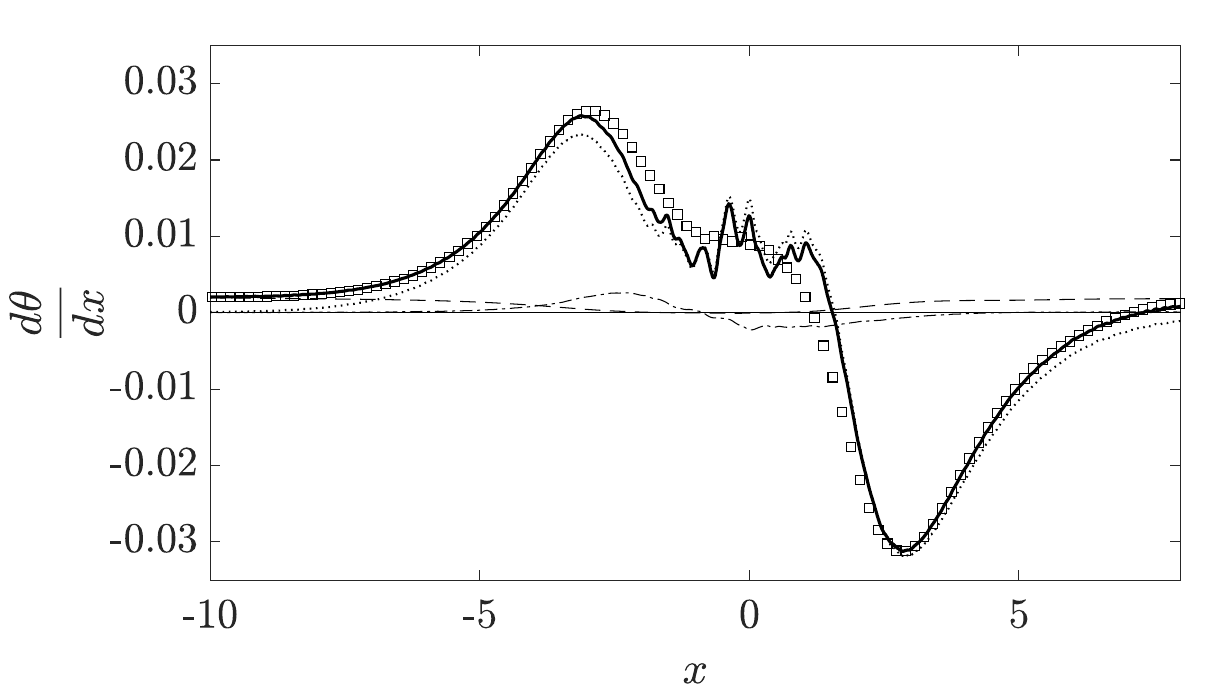}
    \caption{}
\end{subfigure}
~
\begin{subfigure}[b]{0.49\textwidth}
    \includegraphics[trim=750 400 130 300,clip,width=\textwidth]{figures/VKI_kawai_C0.pdf}
    \caption{}
\end{subfigure}
\caption{Von Karman integral equation applied to the DNS data (Case C0). ($a$) Validation of  Eq.~(\ref{eq:VKI}). ($b$) Streamwise integration of terms in Eq.~(\ref{eq:VKI}) showing their accumulated effect. Squares, left-hand side; thick solid line, sum of right-hand-side terms; dashed line, term $I$; dotted line, term $II$; dotted-dashed line, term $III$; thin solid line, $d\theta/dx = 0$} \label{fig:VKI_kawai_C0}
\end{figure}

In the present study, Eq.~(\ref{eq:VKI}) is first used to analyze contributions from the RHS terms to the spatial development of $\theta$. This provides an indication of the relative importance of these terms in different regions of the flow. In particular, we are interested in quantifying the role of skin friction in boundary layer development in different nonequilibrium regions (i.e., in upstream APG, downstream FPG, and within the separation bubble). Before using Eq.~(\ref{eq:VKI}) for this analysis, it is important to validate the balance of this equation by evaluating RHS terms from the DNS field data. The left-hand side term is directly evaluated by numerically differentiating $\theta$ obtained from the DNS and is then compared to the sum of RHS terms. Figure \ref{fig:VKI_kawai_C0}($a$) shows this 
validation along with the individual terms on the RHS. The balance is satisfactory except in the close vicinity of the separation bubble. This is likely due to the breakdown of the thin boundary-layer assumption in this region as the boundary-layer thickness ($\delta_{99}$) increases drastically due to separation. It is observed from Fig. \ref{fig:VKI_kawai_C0}($a$) that the skin friction dominates the balance far upstream and downstream of the bubble, with negligible contributions from terms $II$ and $III$. This highlights the need for accurate skin-friction prediction in these zones when we turn to WMLES in section \ref{sec:WMLES}. Term $II$ becomes dominant for $x > -5$ and almost exclusively determines the boundary-layer growth within the bubble. Term $III$ is active only immediately upstream and downstream of the bubble with small positive and negative contributions, respectively. Perhaps the most important observation from this analysis is that the contribution of skin friction to the boundary layer development is negligible in the upstream vicinity of the bubble (starting at $x \approx -5$) and within the bubble. This might suggest that wall modeling (wall-stress modeling, to be precise) in WMLES becomes less relevant beyond $x \approx -5$ for boundary-layer development and separation characteristics. However, through RD decomposition it will be shown in section~\ref{sec:RD_wallmodel} that this is not necessarily true, given the importance of modeling nonequilibrium terms near the wall.

To investigate the accumulated effect of each of the RHS terms on the momentum thickness, Eq.~(\ref{eq:VKI}) can be integrated along the streamwise direction up to an arbitrary $x$. Figure~\ref{fig:VKI_kawai_C0}($b$) shows the result of this integration for the individual terms and their sum. It is observed that the contribution from skin friction plateaus at around $x=-4$,  and becomes active again downstream of the bubble, affirming that skin friction is important outside and away from the bubble. 

Although the foregoing integral analysis provides valuable insights into the relative importance of skin friction in different regions, it is not necessarily effective in isolating the contributions of different physical processes to the skin friction in the flow. In other words, the term $II$ and $d\theta/dx$ together account for various physical effects such as pressure gradient, convection, etc., which is not desirable when this analysis is used as a diagnostic tool for distinguishing various nonequilibrium effects in the flow. 

\subsection{RD decomposition of mean skin friction}\label{sec:RD_flow_phy}

The decomposition technique we employ in this section is called the RD identity proposed by \citet{RD2016} (hereon referred to as RD decomposition). It is the mean streamwise kinetic energy budget integrated across the TBL at each streamwise ($x$) location in the absolute frame of reference (i.e. frame of reference of freestream) where the wall is moving with the velocity $-U_{e}$ ($U_{e}$ being the boundary-layer edge velocity). The rationale behind using the absolute reference frame is that skin friction can be introduced into the kinetic energy budget via the non-zero power imparted by the moving wall to the boundary layer. The RD identity decomposes the mean skin friction into contributions from different physical phenomena in the boundary layer commensurate with their energy contribution. In the following subsections, we first apply the original RD decomposition (in the streamwise direction) to Case C0 and show its validity for the present flow. We then analyze different terms in the RD decomposition to investigate how the energy distribution evolves at various downstream locations. Finally, we use RD decomposition to compare the chordwise component of skin friction from Cases C0 and C35, and extend RD decomposition to the spanwise kinetic energy budget to compare spanwise and chordwise components in Case C35. 


\subsubsection{Case C0: Separation bubble without sweep (2D-TBL)}\label{sec:RD_flow_physics_2D}

Following Eq.~(2.8) of \citet{RD2016}, the RD decomposition for the mean streamwise skin-friction coefficient $C^*_{f,x}$ is given by,

\begin{equation}
    \label{eqn:RD_x_DNS}
     \begin{aligned}
        \frac{C^*_{f,x}}{2} =\; & \overbrace{ \frac{1}{U_{e}^3}\int_{0}^{\delta_{99}} \nu \left( \frac{\partial 	\langle u \rangle }{\partial y} \right)^2 dy }^{C^*_{f1,x}}\; +\; \overbrace{\frac{1}{U_{e}^3}\int_{0}^{\delta_{99}} -\langle u'v' \rangle \frac{\partial \langle u \rangle }{\partial y}  dy }^{C^*_{f2,x}} \\ &+ \underbrace{\frac{1}{U_{e}^3}\int_{0}^{\delta_{99}} \left( \langle u \rangle - U_{e} \right) \frac{\partial}{\partial y}\left( \nu \frac{\partial \langle u \rangle }{\partial y} -\langle u'v' \rangle \right) dy}_{C^*_{f3,x}},
    \end{aligned}
\end{equation}
 
\noindent where $C^*_{f,x}=2\tau_{w,x}/(\rho_e U^2_e)$ is the skin friction based on the local freestream (edge) velocity. Note that the terms $( \nu \frac{\partial \langle u \rangle }{\partial y} -\langle u'v' \rangle )$ in $C^*_{f3,x}$ constitute the total shear stress.
It is worth noting that Eq.~(\ref{eqn:RD_x_DNS}) has been recast in the usual (wall) frame of reference, for practical purposes. However, the physical interpretation of the RHS terms (in the same spirit as \citet{RD2016}) requires a coordinate transformation to the absolute reference frame (see Eq.~(2.6) of \cite{RD2016}). In the absolute reference frame, $C^*_{f,x}$ can be interpreted as the mean power supplied by the moving wall to the fluid above it, $C^*_{f1,x}$ as the viscous dissipation of kinetic energy, $C^*_{f2,x}$ as the energy expended for turbulent kinetic energy production, and $C^*_{f3,x}$ as the rate of gain of kinetic energy by the fluid. The equivalent interpretations in the wall reference frame are straightforward, except for $C^*_{f3,x}$, which is interpreted as the spatial growth of the flow by \citet{RD2016}. Note that the particular form of RD decomposition employed in Eq.~(\ref{eqn:RD_x_DNS}) avoids explicit dependence of $C^*_{f3,x}$ on the streamwise derivatives and admits a local (in $x$) decomposition, which is more convenient from the standpoint of data availability of local profiles and their post-processing \cite{RD2016}. However, $C^*_{f3,x}$ can be further decomposed into the following components to elucidate the contributing physical effects further,
\begin{equation}
    \label{eqn:RD_Cfx3}
        \begin{aligned}
        C^*_{f3,x}= & \underbrace{\frac{1}{U_e^3} \int_0^{\delta_{99}}\left(\langle u\rangle-U_e\right)\left(\langle u\rangle \frac{\partial\langle u\rangle}{\partial x}+\langle v\rangle \frac{\partial\langle u\rangle}{\partial y}\right) d y}_{C^*_{f31,x}} \\
        & +\underbrace{\frac{1}{U_e^3} \int_0^{\delta_{99}}-\left(\langle u\rangle-U_e\right)\left(\nu \frac{\partial^2\langle u\rangle}{\partial x^2}-\frac{\partial\left\langle u^{\prime} u^{\prime}\right\rangle}{\partial x}\right) d y}_{C^*_{f32,x}}+\underbrace{\frac{1}{U_e^3} \int_0^{\delta_{99}}\left(\langle u\rangle-U_e\right)\left(\frac{d p / \rho}{d x}\right) d y}_{C^*_{f33,x}},
        \end{aligned}
\end{equation}
where $C^*_{f31,x}$, $C^*_{f32,x}$, and $C^*_{f33,x}$ represent flow convection, streamwise heterogeneity, and pressure gradient, respectively. The decomposition in Eq.~(\ref{eqn:RD_Cfx3}) was used in APG flows by \citet{fan2020} to show that flow convection and pressure gradient are the dominant terms in the decomposition of $C^*_{f3,x}$, with positive and negative contributions, respectively, in APG flows. Furthermore, the effect of convection was shown to be mainly confined to the outer layer in the APG, whereas the effect of pressure gradient was active across the entire boundary layer.

\begin{figure}[t]
\centering
\includegraphics[trim=110 150 150 140,clip,width=0.6\textwidth]{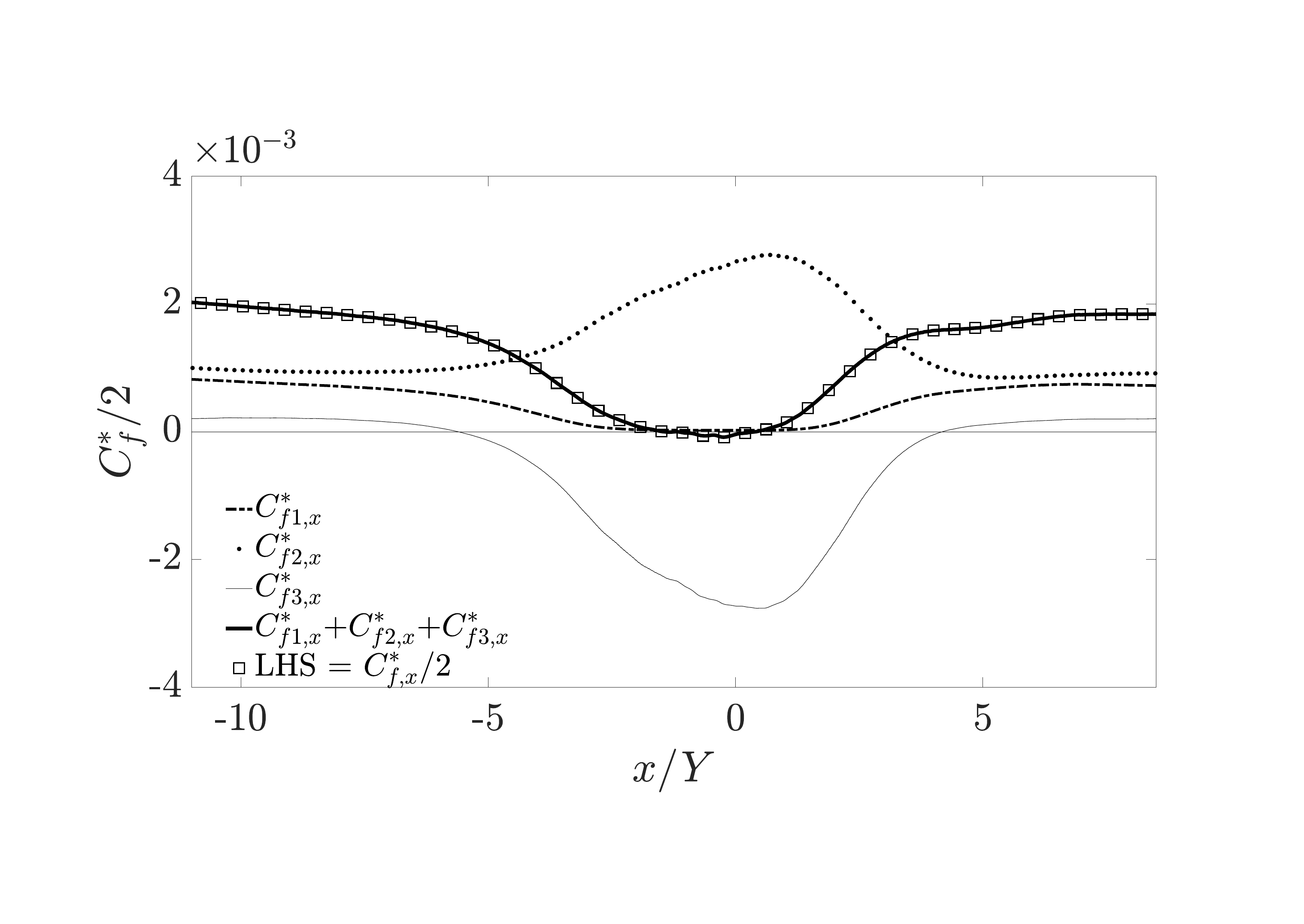}
\caption{Streamwise RD decomposition of skin friction using the DNS data from Case C0. Symbols represent the skin friction data obtained directly from the DNS. Line types represent different components of the RD decomposition in Eq.~(\ref{eqn:RD_x_DNS}) (dotted-dashed, $C^*_{f1,x}$; dotted, $C^*_{f2,x}$; thin solid, $C^*_{f3,x}$; thick solid, sum of the preceding three components).
} \label{fig:RD_Cfx_C0_DNS}
\end{figure}

Figure~\ref{fig:RD_Cfx_C0_DNS} shows the application of the streamwise RD decomposition in Eq.~(\ref{eqn:RD_x_DNS}) to Case C0. The exceptional agreement between the LHS and the sum of RHS terms ascertains the validity of the streamwise RD decomposition in the 2D-TBL. In addition to highlighting the importance of each contributing term in different streamwise (pressure-gradient) zones, the term-wise evolution in $x$ also provides insights into flow physics. In the upstream ZPG region, the contribution from viscous dissipation $C^*_{f1,x}$ is significant and decreases thereafter in the APG until it becomes negligible in the separated region. This contribution is likely concentrated in the inner layer, where the wall-normal gradient of velocity is the steepest and molecular viscosity plays a direct role. Similarly to $C^*_{f1,x}$, the turbulent kinetic energy (TKE) production $C^*_{f2,x}$ shows a significant contribution in the upstream ZPG but unlike viscous dissipation, it keeps increasing through the APG and into the separated region. This is consistent with the expected increase in Reynolds stresses in the detached shear layer as the bubble is approached \cite{Coleman2018}. Note that the decreasing and increasing contributions of $C^*_{f1,x}$ and $C^*_{f2,x}$, respectively, with increasing APG strength were also reported in \cite{fan2020}. The spatial growth term $C^*_{f3,x}$, on the other hand, has a very small contribution in the upstream ZPG likely due to the constant total shear stress in the inner layer, where convection and pressure gradient terms are expected to mostly balance each other \cite{park17aiaa,hickel2012}.
With the increase in APG strength downstream, the pressure gradient dominates the advective term near the wall (as will be explained below through the wall-normal distribution of integrands), resulting in an overall negative contribution, that keeps increasing in magnitude into the separated region. Eventually, the balance between $C^*_{f2,x}$ and $C^*_{f3,x}$ in the separated region determines the characteristics of the thin separation bubble. However, from the wall modeling perspective, only the inner layer contributions are important. Given that most of the TKE production in the $x$-vicinity of the bubble happens away from the wall \cite{Coleman2018} (outside the typical wall-modeled region in WMLES), $C^*_{f3,x}$ seems to be the most crucial term a wall model must capture for accurate prediction in the separated region. This point will become evident in Section~\ref{sec:RD_wallmodel}.

\begin{figure}
\centering
\includegraphics[trim=0 80 0 90,clip,width=\textwidth]{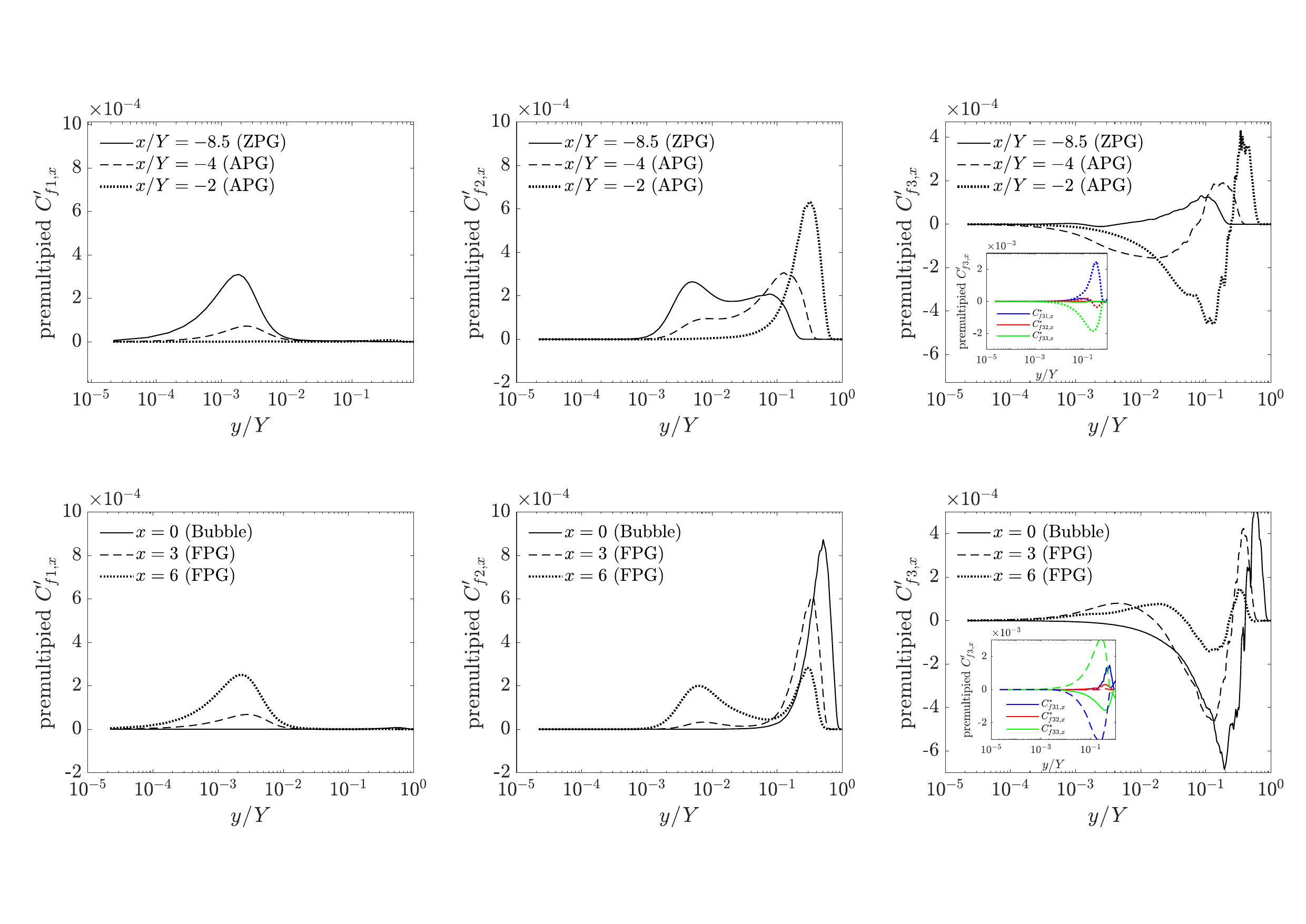}
\caption{Wall-normal distribution of pre-multiplied integrands of components of RD decomposition in Case C0. Top row, upstream of the bubble; bottom row, within and downstream of the bubble. First column, $C'_{f1,x}$; second column, $C'_{f2,x}$; third column, $C'_{f3,x}$. Line types represent different $x$ locations. Insets in the third column show the decomposition of $C^*_{f3,x}$ into sub-constituents as per Eq.~(\ref{eqn:RD_Cfx3}); colors distinguish these sub-constituents (blue, $C'_{f31,x}$; red, $C'_{f32,x}$; green, $C'_{f33,x}$). Line types in insets are the same as the main figures. } \label{fig:premult_RD_C0}
\end{figure}

To get a more precise picture of the wall-normal distribution and redistribution of the contributing terms with the downstream evolution of the TBL, we inspect the integrands (denoted by $C'_{fi,x}$, where $i=1,2,3$) of the RHS terms in Eq.~(\ref{eqn:RD_x_DNS}). In particular, we plot the pre-multiplied integrands ($yC'_{fi,x}$) against $y$ in log scale in Fig.~\ref{fig:premult_RD_C0}, following the analysis of \citet{fan2020}. 
Note that in contrast to the scaling of terms with the local viscous units employed in \cite{fan2020}, we analyze the premultiplied integrands in their dimensional form. The main reason for this choice is the presence of flow separation, which would require either a composite scaling based on both the friction velocity and the pressure gradient \cite{duprat2011} or a purely pressure gradient-based scaling \cite{stratford1959,coleman2017}, in the vicinity of the separation bubble. However, such mixed scaling would obscure a direct comparison of the contributing terms between different streamwise zones and is therefore avoided for simplicity.

The pre-multiplied $C'_{f1,x}$ in Fig.~\ref{fig:premult_RD_C0}(first column) shows that the viscous dissipation is indeed only significant in the inner layer of the upstream ZPG, and keeps decreasing through the APG. This is likely due to the redistribution of energy to the outer layer, as also pointed out by \citet{fan2020}. The redistribution of energy becomes more evident through pre-multiplied $C'_{f2,x}$ in Fig.~\ref{fig:premult_RD_C0}(middle column), where a strong outer peak of TKE production starts to emerge in the APG. In the vicinity of and within the separation bubble ($x$=-2,0 in Fig.~\ref{fig:premult_RD_C0}), the TKE production vanishes in the inner layer and all contribution comes from the outer layer peak, which keeps growing through the separation zone. This lends further weight to the foregoing assertion that for wall models, capturing the TKE production within the modeled region might not be critical for the accurate prediction of skin friction in the separated zone. Downstream of the bubble, recovery of viscous dissipation to the ZPG state is rather quick, with the ZPG profile nearly recovered at $x=6$. However, the TKE production retains a strong footprint of the outer peak at this downstream distance, showing a slower recovery toward the ZPG state. 

Perhaps the most drastic redistribution of energy 
in the wall-normal direction is observed in $C^*_{f3,x}$. Fig.~\ref{fig:premult_RD_C0}(top right) shows that, in the upstream ZPG, the total shear stress is constant (i.e., the $y$-derivative in the integrand of $C^*_{f3,x}$ in Eq.~(\ref{eqn:RD_x_DNS}) is zero) for most of the inner layer, 
and a mild effect of the convection term 
is visible in the outer layer through an overall positive contribution to $C^*_{f3,x}$ (the inset in Fig.~\ref{fig:premult_RD_C0}(top right) shows that convection contributes positively to $C^*_{f3,x}$ in the upstream ZPG). With the imposition of APG, the near-wall balance is dominated by the APG, resulting in a strong negative near-wall contribution. The positive contribution of the convection term in the outer layer also increases at the same time, albeit at a smaller rate than the negative pressure-gradient contribution in the inner layer (see inset in Fig.~\ref{fig:premult_RD_C0}(top right)). This explains the overall negative contribution from $C^*_{f3,x}$ in the APG and the separated zones in Fig.~\ref{fig:RD_Cfx_C0_DNS}. From the wall-modeling perspective, the dominance of near-wall balance by the pressure gradient suggests (similar to observations of \citet{wang02}) that 
directly accounting for only pressure gradient in the wall model is perhaps the most critical for near-wall modeling of nonequilibrium TBLs (see $x$=-2,0 lines in the insets of Fig.~\ref{fig:premult_RD_C0}(third column)). However, depending on the wall-model matching location, the contribution from convection may start to become significant in the wall-model energy balance. 
Regardless, the significant contribution to skin friction from nonequilibrium terms in the near wall is established well from Fig.~\ref{fig:premult_RD_C0}(third column), and any wall model that ignores these terms will poorly predict wall shear stress in nonequilibrium flows (see Section~\ref{sec:RD_wallmodel}). Similarly to the TKE production term, the recovery of $C'_{f3,x}$ profile to the ZPG state downstream of the separation bubble is rather slow, with the upstream history effect visible in the outer profile. The presence of history effects mainly in the outer layer has been noted previously, albeit indirectly, in \cite{hayat2024aiaa,tamaki2020}.

\subsubsection{Case C35: Separation bubble with sweep (3D-TBL) }\label{sec:RD_flow_phy_3D}

\begin{figure}[t]
\centering
\begin{subfigure}[b]{0.54\textwidth}
    \includegraphics[trim=110 150 150 140,clip,width=\textwidth]{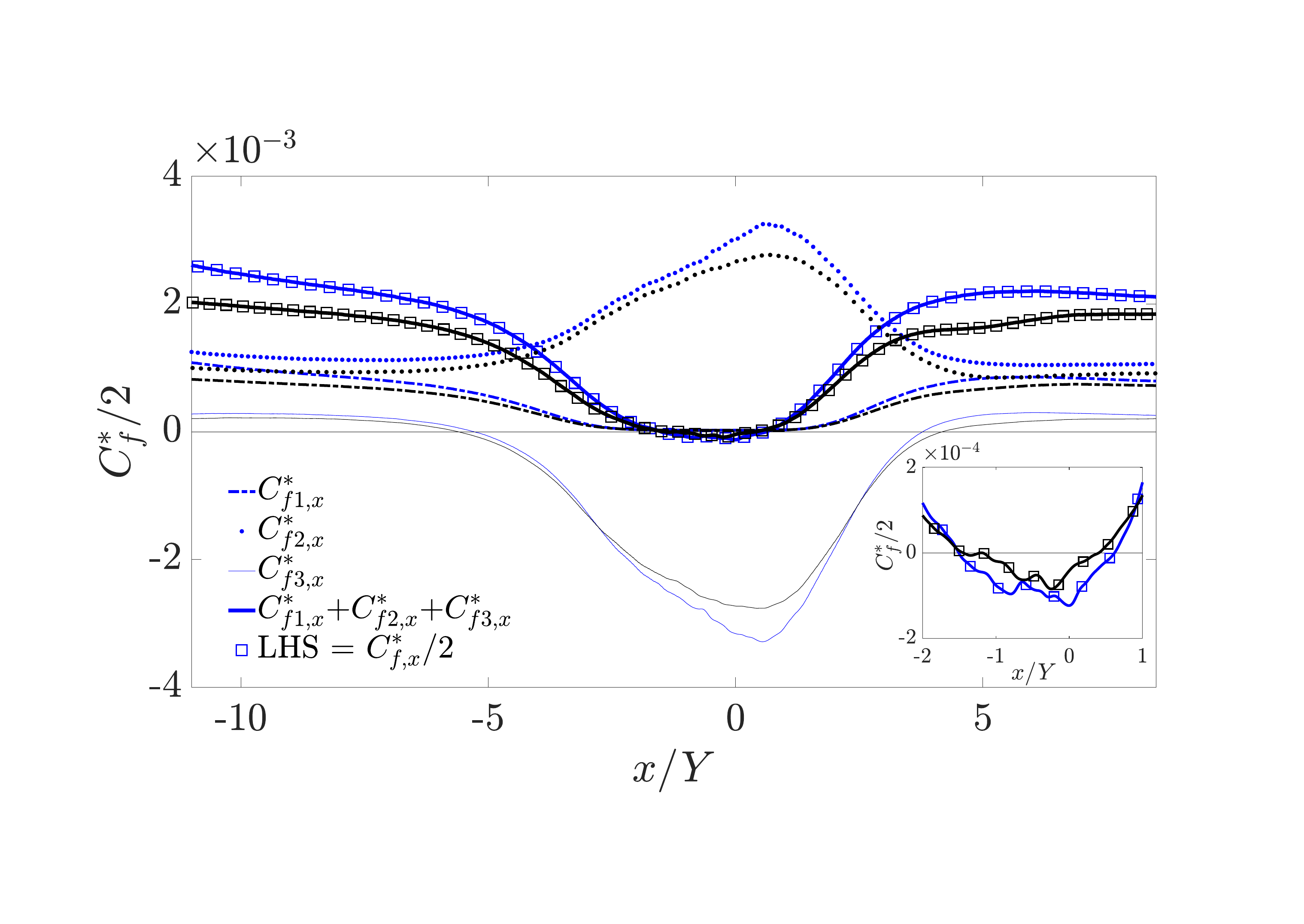}
    \caption{}
\end{subfigure}
~
\begin{subfigure}[b]{0.43\textwidth}
    \includegraphics[trim=90 240 120 250,clip,width=\textwidth]{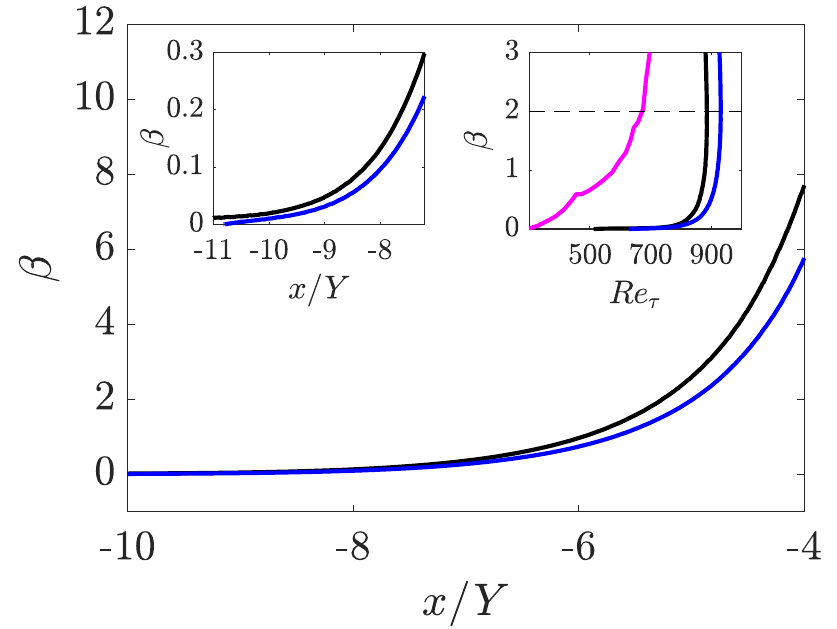}
    \caption{}
\end{subfigure}
\caption{Comparison between C0 and C35 cases, of ($a$) the DNS-based chordwise RD decomposition of skin friction $C^*_{f,x}$ and ($b$) upstream history of $\beta$, as a function of chordwise distance $x$ (and friction Reynolds number in the top-right inset). Colors denote different cases (black, C0; blue, C35; magenta, Case W10 in \cite{fan2020}); line types and symbols in ($a$) are the same as in Fig.~\ref{fig:RD_Cfx_C0_DNS}. Inset in ($a$) magnifies the bubble region to show differences in the skin-friction magnitude. Top-left inset in ($b$) magnifies the upstream region. In the top-right inset of ($b$), the dashed line shows the matching $\beta$ between Case C0 and W10 for comparing integrand profiles in Fig.~\ref{fig:premult_RD_C0_vs_C35_DNS}.}
\label{fig:RD_C0_vs_C35_DNS}
\end{figure}

We first use the streamwise RD decomposition in Eq.~(\ref{eqn:RD_x_DNS}) to compare the chordwise ($x$) components of skin friction from Cases C0 and C35, as shown in Fig.~\ref{fig:RD_C0_vs_C35_DNS}($a$). Two effects are noteworthy from this comparison: first, the skin friction enhances in the upstream ZPG in Case C35, which is attributed to an increase in $C^*_{f1,x}$ and $C^*_{f2,x}$ components in the RD decomposition; 
second, Case C35 has a larger magnitude of negative $C^*_{f,x}$ in the separation bubble, leading to larger bubble size, both in $x$ and $y$ directions (as shown in \cite{Coleman2019}).

\begin{figure}
\centering
\includegraphics[trim=0 80 0 90,clip,width=\textwidth]{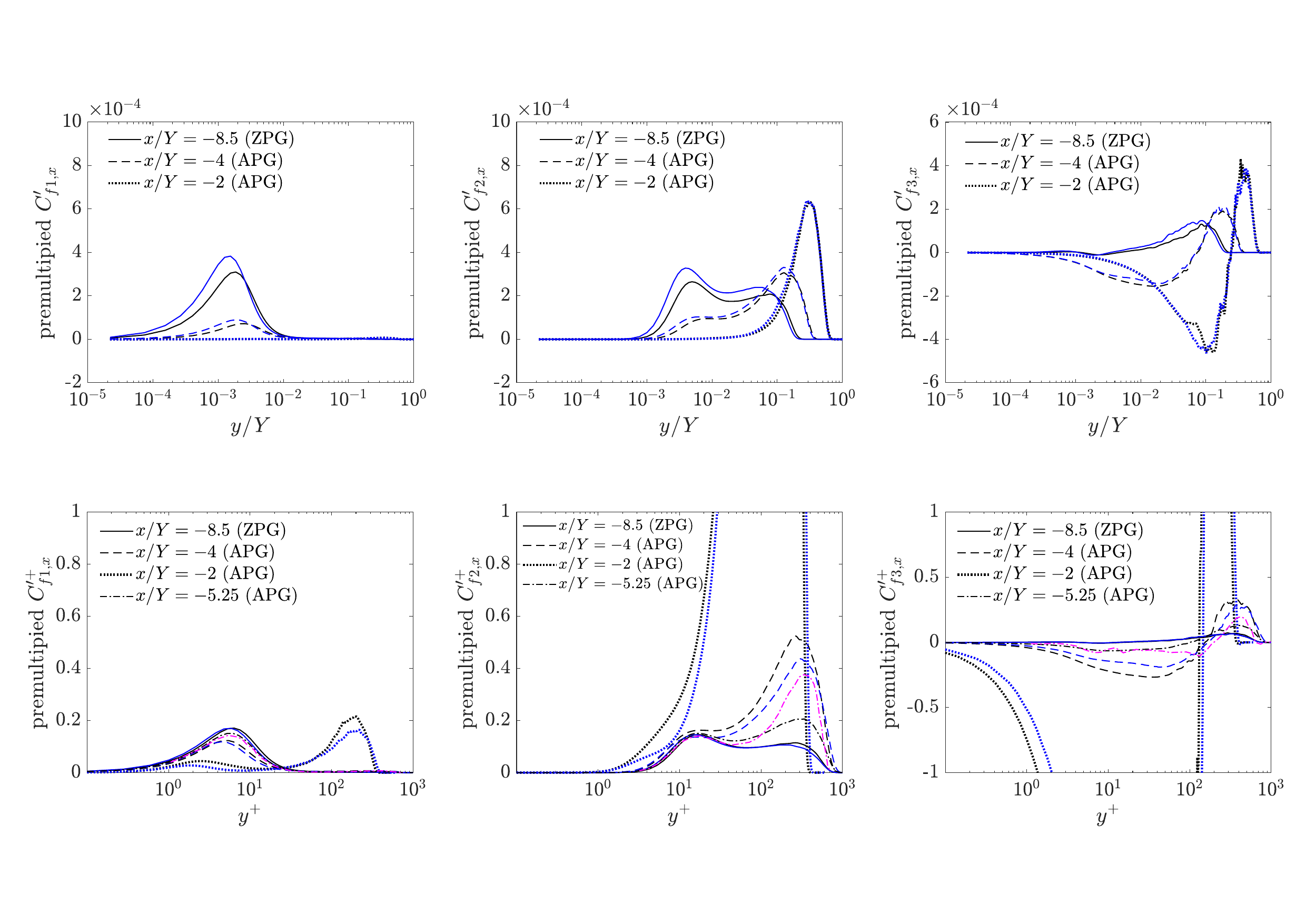}
\caption{Comparison of pre-multiplied integrands of components of RD decomposition between C0 and C35 cases, using nominal scaling (top row) and viscous wall scaling (bottom row). Black, Case C0; blue, Case C35; magenta, Case W10 in \cite{fan2020} at $\beta \approx 2$. Line types represent different $x$ locations. } \label{fig:premult_RD_C0_vs_C35_DNS}
\end{figure}

A physical explanation of the foregoing effects is based on the Clauser pressure-gradient parameter $\beta=\delta_x^*/\tau_{w,x} dP/dx$ and its upstream development history. Figure~\ref{fig:RD_C0_vs_C35_DNS}($b$) shows the spatial history of $\beta$ upstream of the bubble. Case C0 shows a larger accumulation of $\beta$ history (area under the $\beta \text{-} x$ curve),
which keeps growing with $x$. Note from the top-left inset that $\beta$ is larger for Case C0 even very close to the inflow. As observed by \citet{fan2020}, a stronger accumulated $\beta$ leads to a lower skin fiction, which explains the consistently lower $C^*_{f,x}$ for Case C0 in the upstream ZPG. Furthermore, \citet{fan2020} reported that a larger accumulation of upstream $\beta$ leads to a suppression of the inner-layer dynamics and an enhancement of the outer-layer contributions. Figure~\ref{fig:premult_RD_C0_vs_C35_DNS} (top row) corroborates this effect for the present flow, where suppression of inner peak is observed for the pre-multiplied viscous dissipation $C'_{f1,x}$ and pre-multiplied TKE production $C'_{f2,x}$, at all upstream stations. This explains the reduced $C^*_{f1,x}$ and $C^*_{f2,x}$ in the upstream ZPG for Case C0 in Fig.~\ref{fig:RD_C0_vs_C35_DNS}($a$), as the energy is mainly concentrated in the inner layer in the ZPG. The enhancement of the outer peaks 
is observed more clearly through the pre-multiplied $C'^+_{f2,x}$ in the bottom row of Fig.~\ref{fig:premult_RD_C0_vs_C35_DNS}, where quantities are scaled with inner units. Also included are the profiles for NACA4412 airfoil (called W10 in \cite{fan2020}) at $\beta \approx 2$. For a fair comparison with W10, profiles from Case C0 are plotted at $x/Y=-5.25$, where $\beta \approx 2$ (see Fig~\ref{fig:RD_C0_vs_C35_DNS}($b$)). However, note (from the top-right inset of Fig~\ref{fig:RD_C0_vs_C35_DNS}($b$)) that although $\beta$ is matched, W10 has a larger accumulation of $\beta$ than Case C0, and therefore the outer peak of $C'^+_{f2,x}$ is stronger for W10. 

The energization of large-scale structures in the outer flow, associated with stronger $\beta$ (or APG \cite{harun2013,bradshaw1967,lee2017}) could be the reason for the smaller separation bubble size in Case C0. As pointed out by \citet{tamaki2020}, the outer-layer turbulence dominates the boundary layer development near separation. The more energetic outer large-scale motions in Case C0 likely energize the detaching shear layer, causing a delay in separation (from $x/Y=-1.51$ in C35 to $x/Y=-1.43$ in C0) and effecting a weaker bubble with a smaller height (maximum bubble height: $0.053Y$ in C0 and $0.095Y$ in C35). Finally, from Fig.~\ref{fig:premult_RD_C0_vs_C35_DNS} (bottom-right subplot), the stronger history effect of $\beta$ causes noticeable modifications to the wall-normal distribution of $C'_{f3,x}$ in Case C0. Importantly, the near-wall distribution is affected, which could potentially impact the predictions of wall models that do not incorporate history effects through the growth term $C^*_{f3,x}$. Note that such models nevertheless retain some information on the nonequilibrium terms, such as the local pressure gradient, that is tacitly contained in the velocity at the matching location, as suggested by \citet{larsson16}. However, contrary to their argument of convection and pressure gradient balancing each other in the log layer, $C'_{f3,x}$ distribution in Fig.~\ref{fig:premult_RD_C0_vs_C35_DNS}(bottom-right) shows a strong imbalance of these terms in the log layer of the present flow. Therefore, the upstream accumulated effect of this imbalance coming exclusively from the log layer, if not directly resolved in the wall model, is expected to impact the near-wall flow prediction in the downstream region. Also note from Fig.~\ref{fig:premult_RD_C0_vs_C35_DNS} (bottom-right) that the stronger $\beta$ history in Case C0 enhances the aforementioned imbalance between nonequilibrium forces in the log layer. This is perhaps one reason why lower complexity models like ODE equilibrium and integral nonequilibrium wall models struggle to predict separation accurately in Case C0, as will become evident in Section~\ref{sec:2D_bubble}. 

\begin{figure}[t]
\centering
\includegraphics[trim=110 150 150 140,clip,width=0.6\textwidth]{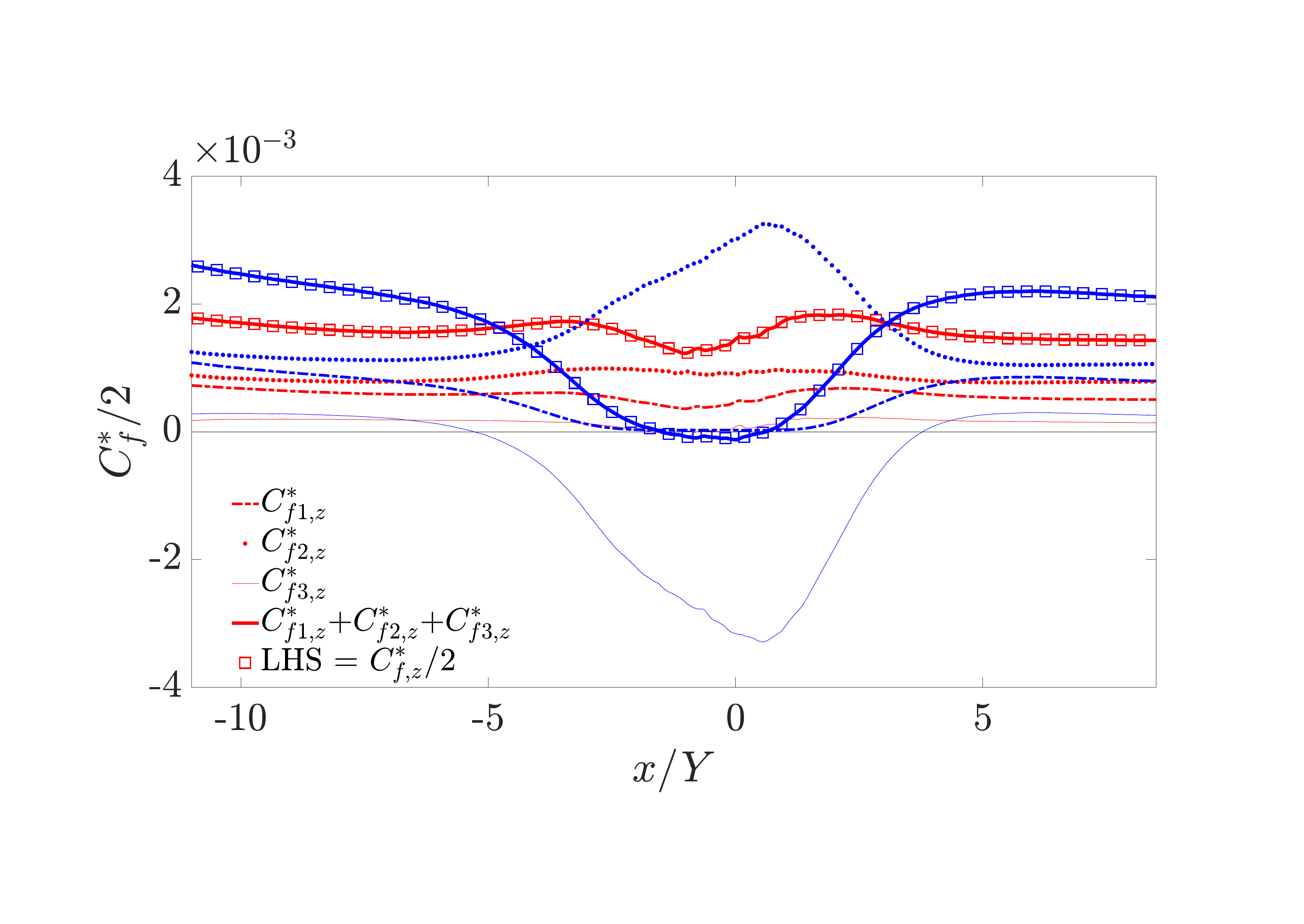}
\caption{Comparison of the RD decomposition of chordwise and spanwise components of skin friction using the DNS data from Case C35. Colors denote RD decomposition in different directions (blue, chordwise; red, spanwise); line types and symbols are the same as in Fig.~\ref{fig:RD_Cfx_C0_DNS}.} \label{fig:RD_Cfx_vs_Cfz_C35_DNS}
\end{figure}

Next, we extend RD decomposition to the spanwise ($z$) component of skin friction. The derivation of spanwise RD decomposition proceeds in the same way as the chordwise one given in Eq.~(\ref{eqn:RD_x_DNS}), except that the chordwise velocity component $u$ is replaced with the spanwise component $w$.
\begin{equation}
    \label{eqn:RD_z_DNS}
     \begin{aligned}
        \frac{C^*_{f,z}}{2} =\; & \underbrace{ \frac{1}{W_{e} U_{e}^2}\int_{0}^{\delta_{99}} \nu \left( \frac{\partial 	\langle w \rangle }{\partial y} \right)^2 dy }_{C^*_{f1,z}}\; +\; \underbrace{\frac{1}{W_{e} U_{e}^2}\int_{0}^{\delta_{99}} -\langle w'v' \rangle \frac{\partial \langle w \rangle }{\partial y}  dy }_{C^*_{f2,z}} \\ &+ \underbrace{\frac{1}{W_{e} U_{e}^2}\int_{0}^{\delta_{99}} \left( \langle w \rangle - W_{e} \right) \frac{\partial}{\partial y}\left( \nu \frac{\partial \langle w \rangle }{\partial y} -\langle w'v' \rangle \right) dy}_{C^*_{f3,z}},
    \end{aligned}
\end{equation}
\noindent where $W_e$ is the spanwise component of the local edge velocity and $C^*_{f,z} = 2\tau_{w,z}/(\rho_e U^2_e)$. Note that $U_e^2$ in the denominator stems from our choice of defining the skin friction based on the chordwise freestream velocity (to be consistent with \cite{Coleman2019}). Figure~\ref{fig:RD_Cfx_vs_Cfz_C35_DNS} compares the chordwise ($x$) distribution of $x$ and $z$ components of the RD decomposition in Case C35. It is observed that all $C^*_{f,z}$ components exhibit relatively small variations in $x$ compared to $C^*_{f,x}$ components. Crucially, the relative contribution from each component remains approximately the same, indicating that the spanwise boundary layer largely retains the characteristics of the upstream ZPG layer. It is also evident that unlike the streamwise spatial growth term $C^*_{f3,x}$, the spanwise growth term $C^*_{f3,z}$ does not dominate the balance in any of the APG, separated, and FPG zones, revealing that nonequilibrium effects (due to pressure gradient and separation) have an insignificant effect on the spanwise velocity profiles. On the other hand, viscous dissipation and production remain the dominant terms (similar to ZPG-TBL) at all chordwise stations. The foregoing observations have an important implication for near-wall modeling: since the energy balance of $W$ profiles at all $x$ approximately matches the upstream ZPG-TBL balance, the log law of the wall likely remains a valid assumption for the spanwise component in the present flow, and even the simple wall models are expected to give reasonable predictions for $C^*_{f,z}$. This will be demonstrated in Section~\ref{sec:3D_bubble}.

\begin{figure}[t]
\centering
\includegraphics[trim=0 80 0 90,clip,width=\textwidth]{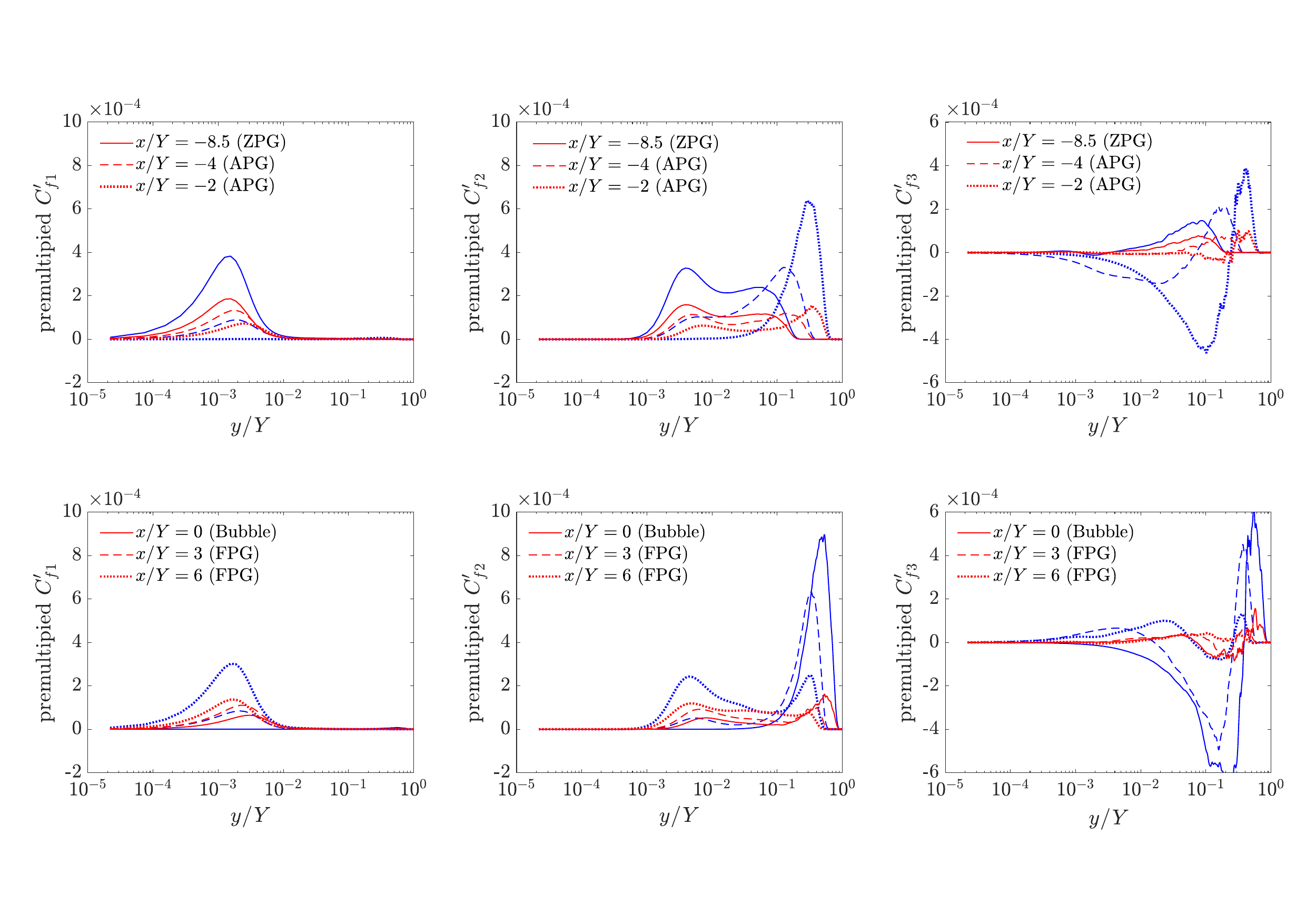}
\caption{Comparison of pre-multiplied components of RD decomposition between chordwise $x$ and spanwise $z$ components of skin friction in Case C35. Colors denote RD decomposition in different directions (blue, chordwise; red, spanwise). Line types represent different $x$ locations.}
\label{fig:premult_Cfx_vs_Cfz_RD_DNS_C35}
\end{figure}

To further consolidate the above conjecture and contrast the modeling requirements for spanwise vs. chordwise $C^*_{f}$ components, we analyze the wall-normal distribution of RD components in Fig.~\ref{fig:premult_Cfx_vs_Cfz_RD_DNS_C35} for Case C35. Firstly, $C'_{f1,z}$ remains non-zero within the separation bubble. Secondly, and more importantly, the inner peak of $C'_{f2,z}$ persists throughout the APG and separated zones, indicating that spanwise flow remains sufficiently turbulent close to the wall. This is in stark contrast to $C'_{f2,x}$, where the inner layer turbulence vanishes in the separated zone. Both these observations on $C'_{f1,z}$ and $C'_{f2,z}$ emphasize the need for wall modeling in the spanwise direction within the separated zone. Based on mostly 2D-TBLs, past studies have suggested that wall modeling may be unnecessary in the upstream vicinity of and within the separation bubble \cite{tamaki2020} and others propose turning off the wall model within the separation zone \cite{bodart2013}. The present analysis shows that wall modeling remains critical for all zones in 3D-TBLs, which in practical flows are the norm rather than the exception. The outer layer distribution of $C'_{f2,z}$ sheds further light on the differing physics of spanwise and chordwise components. Unlike chordwise production, whose outer peak dramatically increases through APG and separated zones, spanwise production does not show an appreciable increase in the outer peak. This indicates that upstream of, within, and downstream of the separated zone, the spanwise flow is largely decoupled from the outer-layer energization in the chordwise direction caused by the pressure gradient. This is consistent with the validity of the independence principle in this region, as pointed out by \citet{Coleman2019}. Finally, $C'_{f3,z}$ distribution shows only a weak chordwise evolution of the outer layer of spanwise TBL, lending further weight to the absence of nonequilibrium effects in the spanwise component. Crucially, from the wall modeling perspective, the inner layer ($y/Y$ up to $\sim$0.03) of the spanwise TBL remains mostly unaffected by the pressure gradient, suggesting that modeling nonequilibrium terms within a wall model is insignificant for the prediction of spanwise skin friction.

Having established above, the utility of RD decomposition in analyzing physically distinct contributions to skin friction, we will now extend this analysis framework to wall models in Section \ref{sec:RD_wallmodel}. The goal is to identify, by direct comparison with the DNS results, physical terms in each wall model that contribute to discrepancies in the predicted skin friction coefficient.

\section{Analysis of nonequilibrium effects in WMLES}\label{sec:WMLES}

In this section, we investigate WMLES results for the same reference DNS configurations (\cite{Coleman2018,Coleman2019}) analyzed in Section~\ref{sec:DNS_analysis}. First, we compare WMLES results from different wall models with the DNS, using the overall mean-flow profiles from the LES grid. The differences in wall-model results are explained through apparent/implementation aspects of wall models as well as based on the physical insights gained from analyzing the DNS datasets in Section~\ref{sec:RD_flow_phy}. Furthermore, the outer-layer results are analyzed based on certain physical aspects such as the skewing mechanism. Finally, the RD decomposition framework from Section~\ref{sec:RD_flow_phy} is applied specifically to the wall-modeled region (using wall-model equations), to reveal the differences in their underlying physical mechanisms. 

A few remarks are warranted about the challenges the reference DNS cases present to the WMLES. First, the Reynolds number is sufficiently high (compared to earlier DNSs of the same configuration \cite{na1998,spalart1997}) to make the wall-stress modeling relevant. Two features of the present flow make it especially challenging for WMLES. First, the separation is pressure-gradient-induced, which makes it relatively difficult to capture accurately compared to (massive) separation induced by an abrupt change in geometry, where separation is almost guaranteed to occur. To cause the boundary layer to separate, the upstream distribution of pressure-gradient, free-stream velocity, and boundary-layer integral quantities must be matched precisely with the DNS. Secondly, the separation bubble  is extremely thin 
in its height 
in Case C0, making it a limiting case for wall models to predict. Indeed, it will become evident in section \ref{sec:2D_bubble} that the less flexible wall models (with built-in assumptions about profiles) fail to predict reversed wall shear stress unless provided with the right set of conditions at the matching location. The primary challenge in this configuration is predicting the mean flow within the  FPG and APG regions, where the equilibrium assumptions are severely violated. Furthermore, the prediction of the separation bubble size characteristics such as its streamwise extent and height, and the location of mean separation and reattachment, is of particular interest in the present study. Our goal is to deploy wall models of increasing complexity in this flow and to investigate if the direct incorporation of nonequilibrium effects in more complex wall models improves the WMLES predictions in the nonequilibrium zones and within the separated region. Additionally, in Case C35, we are interested in assessing how faithfully each wall model represents the three-dimensionality of TBL and what impact this would have on the wall-model predictions.

\subsection{Numerical setup of WMLES}\label{sec:comp_setup}

The governing equations for LES are solved using CharLES, an unstructured-grid finite-volume compressible solver developed by Cascade Technologies (Cadence Inc.). An explicit third-order Runge-Kutta (RK3) scheme is used for time advancement and a second-order central scheme is used for spatial discretization. More details regarding the flow solver can be found in \cite{park16jcp}. The constant-coefficient Vreman model \citep{vreman1994} is used for the closure of the SGS stress tensor.

\begin{figure}[t]
\centering
\begin{subfigure}[b]{0.54\textwidth}
    \includegraphics[trim=0 100 450 100,clip,width=\textwidth]{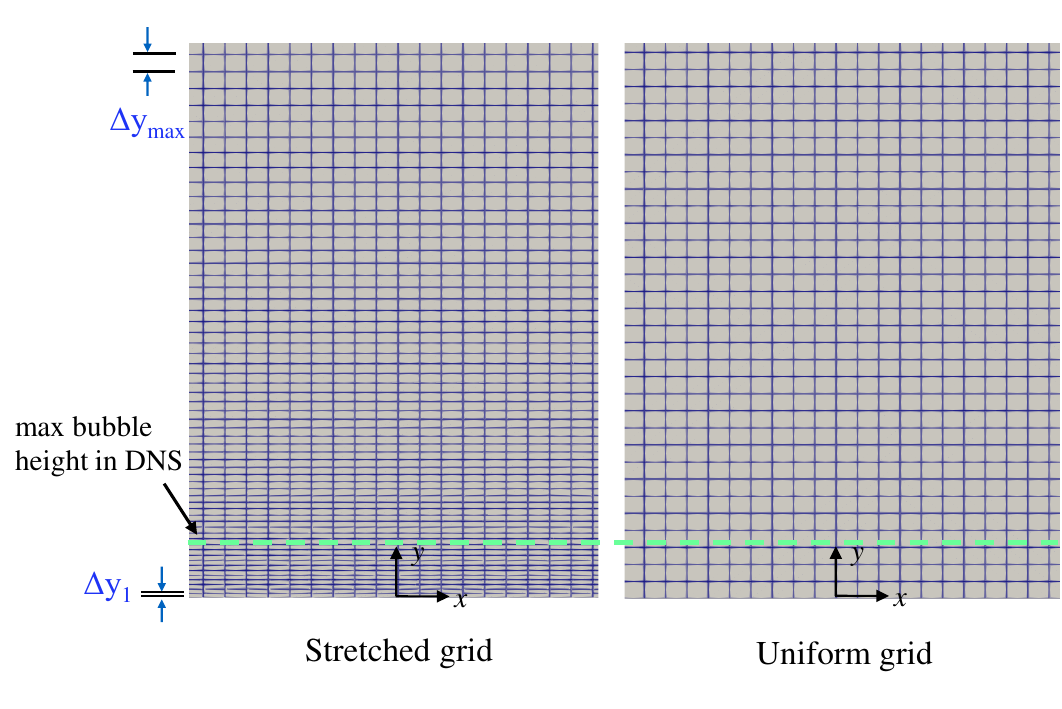}
    \caption{}
\end{subfigure}
~
\begin{subfigure}[b]{0.42\textwidth}
    \includegraphics[trim=50 240 90 250,clip,width=\textwidth]{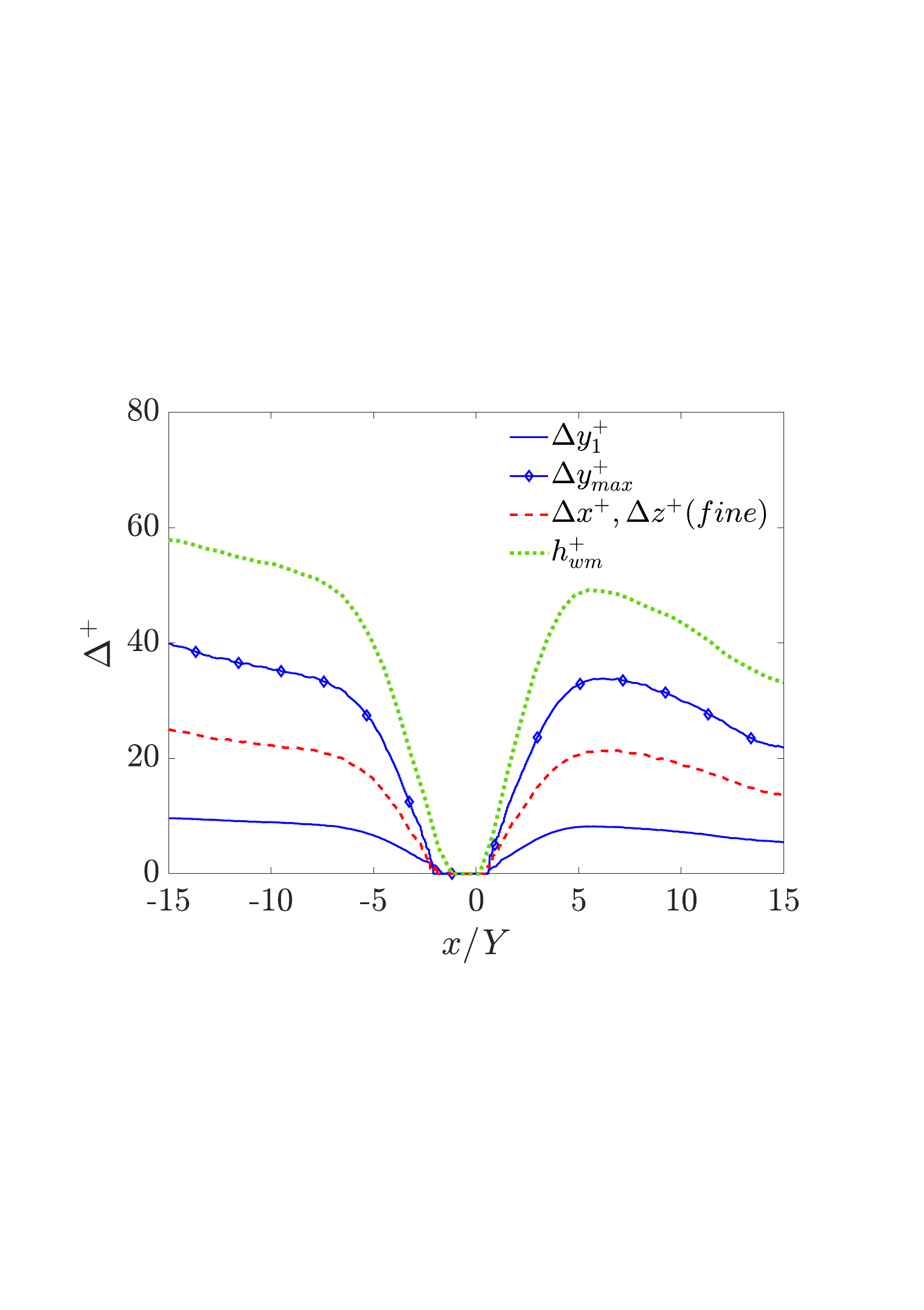}
    \caption{}
\end{subfigure}
\caption{Grid information for WMLES of 
Case C0 and C35. (a) Grid types: stretched in $y$ (left) and uniform in $y$ (right).  (b) $x$-distribution of grid spacing and wall-model matching height in wall units. The green dashed line in ($a$) shows the maximum bubble height for Case C0, to facilitate a comparison of the number of grid points covered by the maximum bubble height in each grid type.
} \label{fig:grid_sepbub}
\end{figure}

A Cartesian rectangular grid is used in the simulations, with uniform chordwise ($x$) and spanwise ($z$) spacing. The grid is stretched (non-uniform) in the wall-normal direction ($y$), with grid spacing increasing away from the wall, as shown in Fig.~\ref{fig:grid_sepbub}$(a)$. This choice for $y$-spacing was driven by the small height of the separation bubble. Particularly, in Case C0, the maximum bubble height is so small that a uniform $y$-spacing with a typical WMLES resolution ($\Delta y^+ \approx 40$ in the ZPG region) results in only $1-2$ cells across the local height of the bubble at most of the streamwise locations, as depicted in Fig.~\ref{fig:grid_sepbub}$(a)$. With the matching location placed at the third cell or higher to avoid log-layer mismatch \cite{kawai12}, this meant that the wall model would not be fed the reversed flow information at the matching location. This point is particularly problematic for lower fidelity wall models incapable of representing sign change in the wall-model velocity profile. This limitation will become more evident in section \ref{sec:2D_bubble}. 
Simulations for both cases were performed at two levels of grid refinement. For the fine grid, grid spacing in each wall-parallel direction was refined twice compared to the coarse grid, resulting in $15$ million and $60$ million grid points for the coarse and fine grids, respectively. However, results from only fine-grid simulations are shown in the following sections. The coarse-grid simulations were only used to establish grid convergence (not shown). Figure~\ref{fig:grid_sepbub}$(b)$ shows grid spacings in wall units for coarse and fine grid simulations, whereas information about grid spacing in terms of domain height ($Y$) is provided in Table \ref{tab:table2}. Hexahedral cells were used throughout the domain. The simulation time step for C0 and C35 cases were respectively set to $\Delta t^{+} = \Delta t \: u^{2}_{\tau,\text{ZPG}}/\nu \approx 0.084$ and $0.157$, where $u_{\tau,\text{ZPG}}$ is the magnitude of total friction velocity at the reference ZPG location in each case.

\begin{table}[t]
\caption{\label{tab:table2} Computational grid information. Note that all grid spacings are given as a fraction of maximum domain height $Y$. $\Delta y_{1}$ and $\Delta y_{max}$ are the minimum (or first off-wall cell) and maximum wall-normal grid spacing, respectively; $\Delta x$ and $\Delta z$ are the wall-parallel grid spacing.}
\centering
\begin{tabular}{lccc}
\hline
\hline

Grid & $\Delta y_{1}$ & $\Delta y_{max}$ &  $\Delta x = \Delta z$ \\\hline
Coarse  & 0.0025 & 0.01 & 0.0125    \\
Fine    & 0.0025 & 0.01 & 0.00625   \\
\hline
\hline
\end{tabular}
\end{table}

In the reference DNS, the domain was semi-infinite in the wall-normal ($y$) direction, courtesy of the spectral code used therein, and the transpiration profile in Eq.~(\ref{eqn:transpiration}) was imposed at a virtual surface at $y=Y$. In the present WMLES, we employ a finite wall-normal domain of height $Y$ and impose the following irrotational transpiration boundary condition at the top boundary of the domain (as suggested by \citet{Coleman2018} in Eq.~$2.2(a-c)$),
\begin{equation}
    \label{eqn:top_bc}
        v = V_{top}(x),~\; \frac{\partial u}{\partial y} = \frac{\text{d}V_{top}}{\text{d}x},~\; \frac{\partial w}{\partial y} = 0.
\end{equation}
Both the spanwise ($z$) periodic boundary condition and the spanwise domain extent of the DNS are replicated in the present study. In the streamwise ($x$) direction, a longer domain than the DNS was required in the present study to allow the development of ZPG-TBL at the inflow, as well as to impose a sponge region at the outflow, allowing the boundary layer to exit the domain smoothly without introducing spurious oscillations. The extent of the inflow development and the outflow sponge region are depicted in Fig.~\ref{fig:schematic_sepbub} and their values are given in Table~\ref{tab:table1}. The sponge region terminates in a subsonic outlet Navier-Stokes characteristic boundary condition \citep{poinsot1992}. Note that in the DNS a fringe boundary condition was used at inflow and outflow for recycling the outgoing profile \cite{spalart1993}. This method was specifically developed for the spectral code
used in the DNS (which requires periodicity in $x$) and is not replicated in the present WMLES, instead utilizing a general-purpose finite-volume code.
The change of boundary condition in the WMLES possibly has some implications for the recovery of skin friction downstream of the bubble in Case C0, as will become evident in section \ref{sec:2D_bubble}. Further information on the inflow characterization, the correct imposition of pressure gradient, and outer flow characteristic thicknesses is provided in Appendix~\ref{appendix:inlet_potential_flow}.

An adiabatic wall model is applied to the bottom wall, which imposes a Neumann boundary condition for velocity and temperature at the wall. Three wall models are considered in the present study, namely, an ODE equilibrium stress model (EQWM), an integral nonequilibrium wall model (integral NEQWM), and a PDE nonequilibrium wall model (PDE NEQWM). All these wall models have been tested extensively in CharLES for various equilibrium and nonequilibrium flows \citep{park14pof,park16jcp,park16prf,park17aiaa,bodart2012sensor,hu23,hayat2024aiaa,hayat2023}. A brief description of each wall model is provided in 
Appendix~\ref{appendix:WM_formulations}. For further detail, the reader is referred to \cite{hu23} and \cite{hayat2024aiaa}. The wall models take the LES state as input at their top boundaries and provide the wall shear stress and heat flux as outputs, which are imposed as a Neumann boundary condition at the wall in the LES. In the present study, the matching height $h_{wm}$, at which LES states are fed to the wall model, is fixed at $h_{wm}/Y = 0.015$ for both C0 and C35 cases. This roughly translates to up to 10\% of the local boundary thickness $\delta_{995}$, which is the typical matching location for WMLES. The corresponding matching height in wall units is shown in Fig. \ref{fig:grid_sepbub}$(b)$. 

Mean flow and turbulence statistics reported in this study were probed along the spanwise centerline ($z/Y=0$) and averaged in $z$ and time $t$. All simulations were allowed to run for at least 2 convective flow-through times ($(x_{\rm sponge} - x_{\rm ref})/U_{\infty}$) before the accumulation of statistics was started. Turbulence statistics were averaged in time for at least 2 flow-through times (similar to the DNS), which corresponds to $T^{+} = T u^{2}_{\tau,\text{ZPG}}/\nu  \approx 5900$ and $11600$ wall time scales for C0 and C35 cases, respectively.

\subsection{Mean flow fields from wall-modeled LES}\label{sec:wmles_results}
The analysis in this section is based mainly on the mean flow and turbulence statistics from the LES grid, to investigate overall differences in the flow fields obtained with different wall models. In contrast, in section \ref{sec:RD_wallmodel} we will mainly use solutions directly from within wall models, to analyze differences in wall-model mechanisms that lead to differences in their predicted wall shear stress.

\subsubsection{Case C0: Separation bubble without sweep (2D-TBL)}\label{sec:2D_bubble}

\begin{figure}[t]
\centering
\begin{subfigure}[b]{0.44\textwidth}
    \includegraphics[trim=0 180 600 200,clip,width=\textwidth]{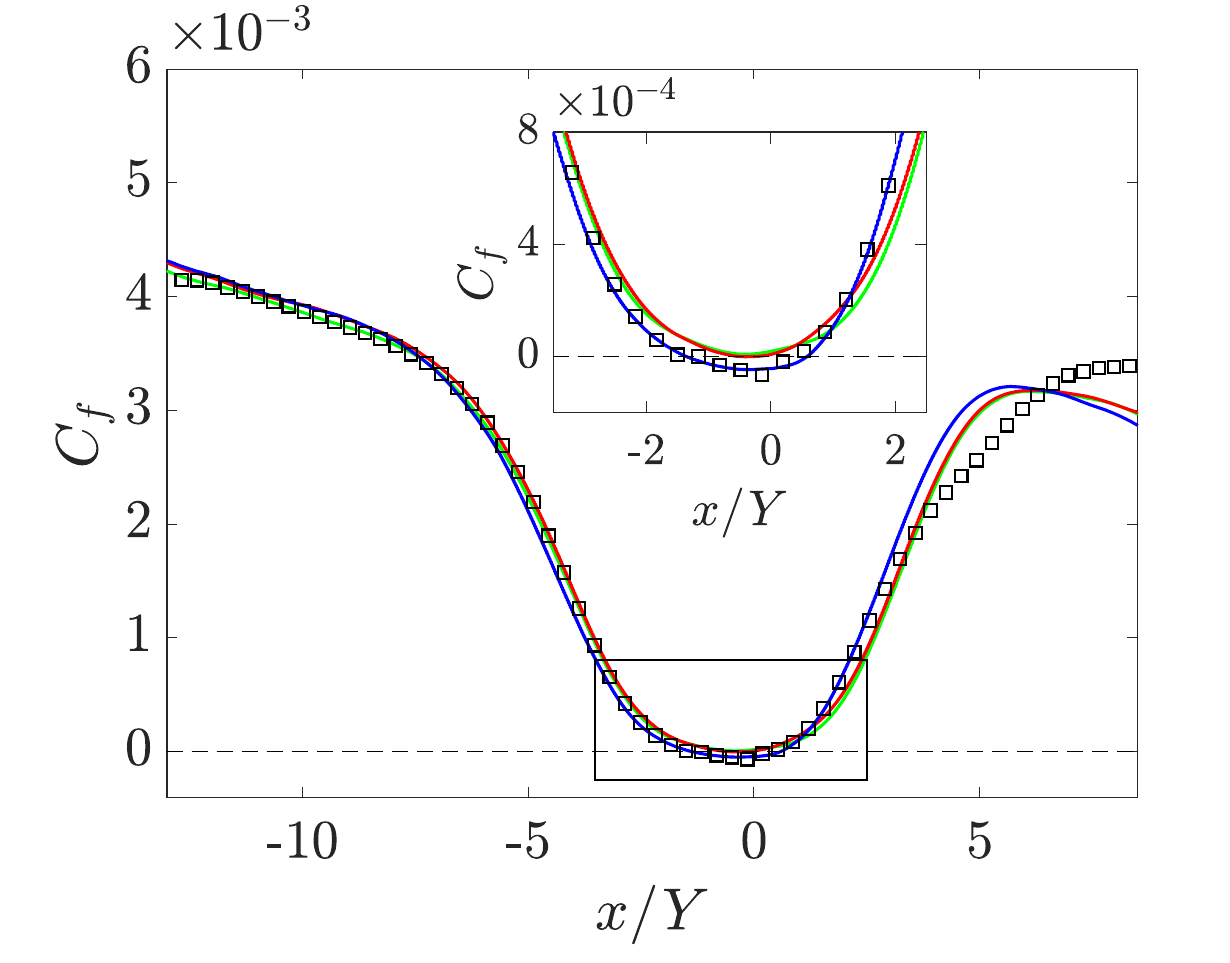}
    \caption{}
\end{subfigure}
~
\begin{subfigure}[b]{0.44\textwidth}
    \includegraphics[trim=590 180 10 200,clip,width=\textwidth]{figures/Cf_Uprofile_C0_WMLES_edit.pdf}
    \caption{}
\end{subfigure}

\caption{$(a)$ Streamwise distribution of mean skin-friction coefficient in Case C0. $(b)$ Mean velocity profiles from LES (solid lines) and wall-model (dotted lines) solutions at $x/Y=0$, to elucidate differences in wall-model mechanisms. Inset in $(a)$ magnifies the boxed region for better visualization of the separated region. Symbols, DNS; red, EQWM; blue, PDE NEQWM; green, integral NEQWM; horizontal dashed line in $(a)$, $C_f=0$; horizontal solid line in $(b)$, wall-model matching height.
} \label{fig:cf_sepbub}
\end{figure}

\begin{figure}[t]
\centering
    \includegraphics[trim=80 380 420 270,clip,width=0.75\textwidth]{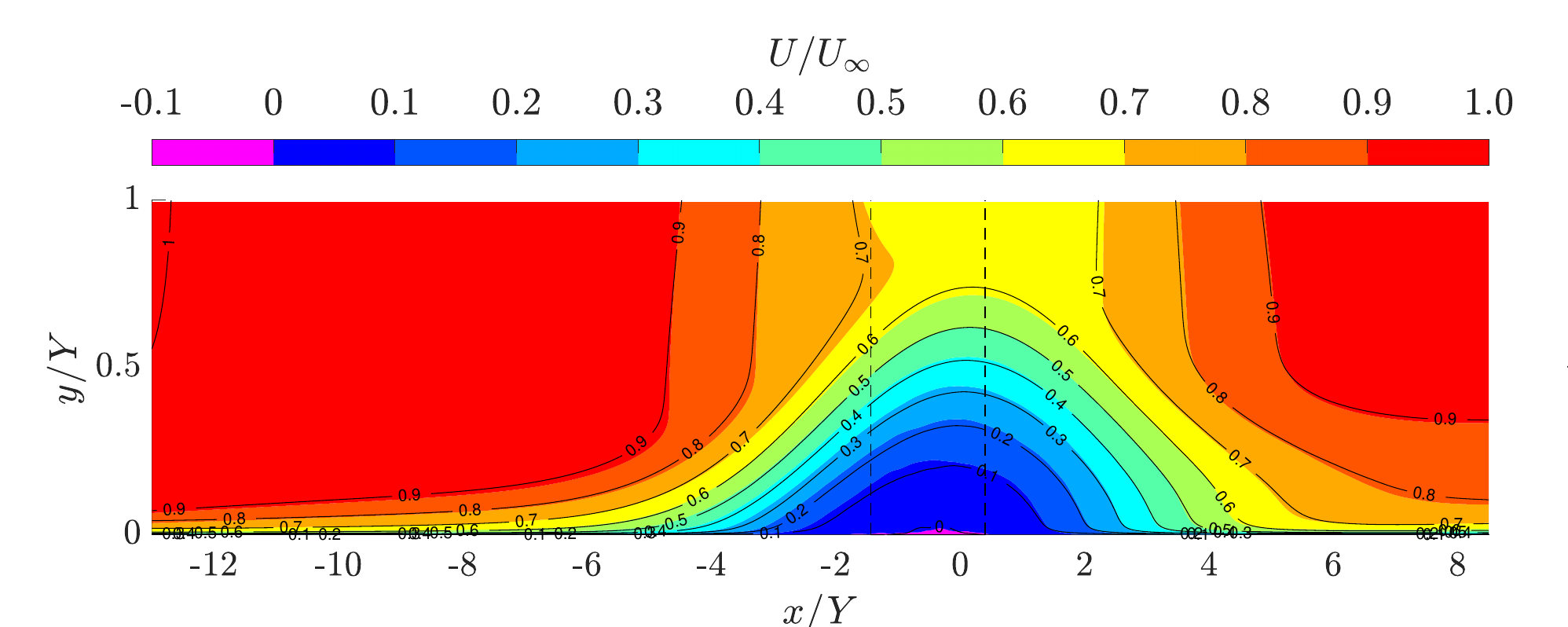}
    
    \includegraphics[trim=80 380 420 270,clip,width=0.75\textwidth]{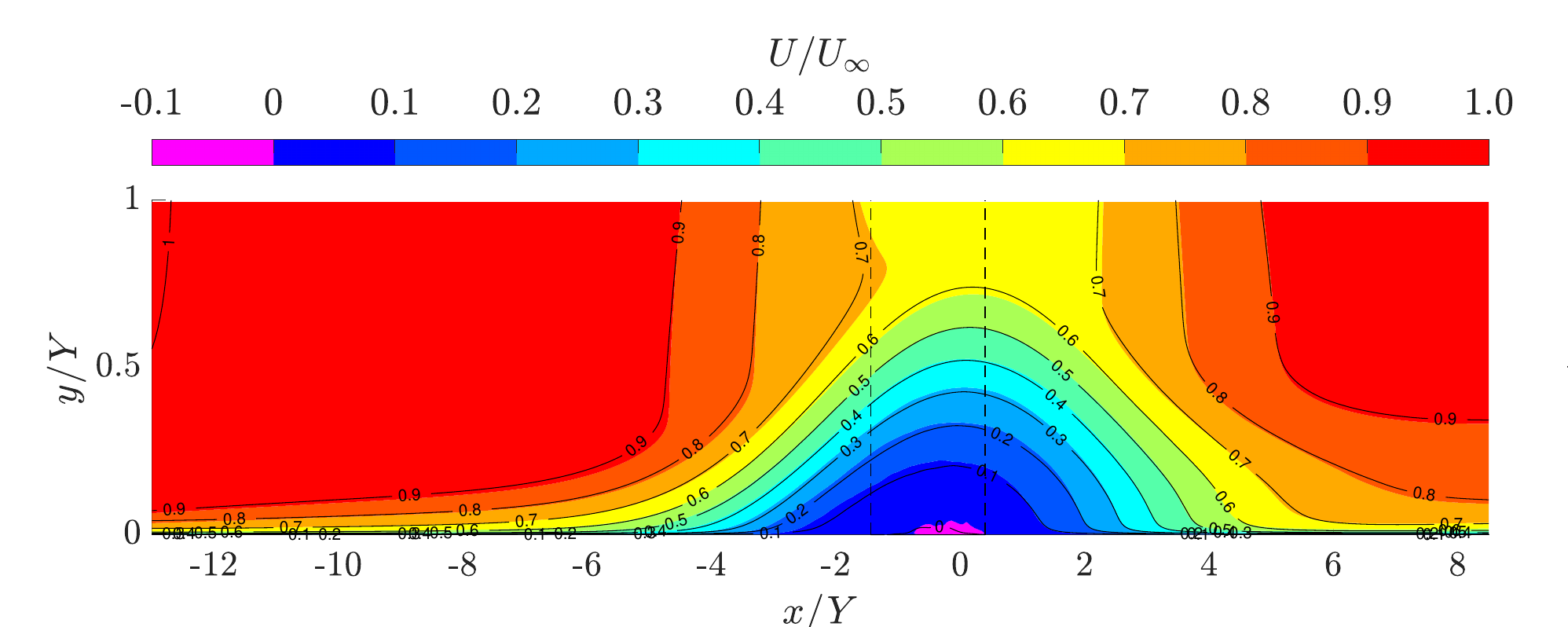}
\caption{Mean streamwise velocity field in Case C0, from the DNS (line contours) and the WMLES (colored contours) using EQWM (top) and PDE NEQWM (bottom). Vertical dashed lines denote $x$ locations of mean separation and reattachment in the DNS.
}
\label{fig:Umean_contours_LES}
\end{figure}

Figure~\ref{fig:cf_sepbub} shows the mean flow quantities from WMLES in Case C0. Note that throughout section~\ref{sec:wmles_results}, skin friction is defined based on the nominal inflow freestream velocity in the $x$ direction, i.e., $C_{f} = 2\tau_w/(\rho_\infty U^2_\infty)$. From Fig.~\ref{fig:cf_sepbub}$(a)$, the predicted $C_{f}$ from all wall models is generally in good agreement with the DNS upstream of the separation bubble, up to the mean separation point. Immediately downstream of reattachment, PDE NEQWM continues to agree with the DNS, whereas EQWM and integral NEQWM deviate appreciably from the DNS, likely due to their inability to predict the bubble (see inset in Fig.~\ref{fig:cf_sepbub}$(a)$). For all wall models, a departure from the DNS is observed for $x/Y> 3$, where \citet{Coleman2018} noted a delayed (linear) recovery of $C_{f}$. Note that $C_{f}$ from the RANS models in \cite{Coleman2018} exhibited a similar departure from the DNS as the present WMLES. Since the RANS simulations also employed a finite wall-normal domain and used the same out-flow treatment as the present WMLES, the departure from the DNS in this region is likely attributed to the difference in the outflow/top boundary conditions between the DNS and WMLES. 
It is evident from the inset in Fig.~\ref{fig:cf_sepbub}$(a)$ that only PDE NEQWM captures the region of negative $C_{f}$, whereas the other two wall models fail to predict any separation (negative $C_f$), at least in the skin-friction plot. However, the mean streamwise velocity field in Fig.~\ref{fig:Umean_contours_LES} depicts a different picture within the separation bubble. First, it is noted in Fig.~\ref{fig:Umean_contours_LES} that both EQWM and PDE NEQWM produce almost similar $U$ fields and show reasonable agreement with the DNS (integral NEQWM results, which are similar to EQWM, are omitted for brevity). More importantly, EQWM LES predicts a region of negative $U$ in the separation bubble. This is in stark contradiction to the $C_{f}$ distribution in Fig.~\ref{fig:cf_sepbub}, which fails to predict a corresponding region of negative $C_{f}$.

To explain this discrepancy between $C_{f}$ and $U$ results, we investigate the solution from within the wall models. Figure~\ref{fig:cf_sepbub}$(b)$ shows the $U$ profile at $x=0$ (located within the separation bubble) from the LES solution and the wall-model solution. Note that although the first off-wall point in the LES grid shows mean flow separation at $x=0$ (in Fig.~\ref{fig:Umean_contours_LES}(top)), the velocity at the matching height is not negative. In other words, EQWM is agnostic to the separation occurring in the few cells next to the wall because, at most $x$ locations, the matching location falls outside the extremely small bubble height in Case C0. This has a strong implication for EQWM and integral NEQWM because, by design, both these wall models do not allow a sign change (flow direction reversal) in the wall-model velocity profile. Therefore, a positive velocity at the matching location predicts a positive $C_f$ for EQWM, as shown in Fig.~\ref{fig:cf_sepbub}$(b)$. 
Furthermore, from Fig.~\ref{fig:Umean_contours_LES}(top), there exists a significant region of separated mean flow (negative $U$ in the first few off-wall cells) that fails to manifest in the wall-model solution due to this limitation. On the other hand, the PDE NEQWM solution is agnostic to such restrictions, owing to its ability to represent sign changes within the wall-model solution, as evident from Fig.~\ref{fig:cf_sepbub}$(b)$. This explains the more accurate prediction of the separated bubble by PDE NEQWM in the inset of Fig.~\ref{fig:cf_sepbub}$(a)$. 

In addition to the aforementioned explanations based on the internal limitations of wall models imposed by their governing equations, it is important to understand the physical processes responsible for the differences in separation predictions. In section~\ref{sec:RD_flow_phy_3D}, it was observed that the stronger $\beta$ history in Case C0 causes an increasing imbalance of nonequilibrium effects in the log layer, as the bubble is approached (see Fig.~\ref{fig:premult_RD_C0_vs_C35_DNS} (bottom-right subplot)). This is directly related to the accumulation of the upstream history effects through the nonequilibrium terms (pressure gradient and convection) in the momentum equation. Since EQWM completely ignores these terms, while the integral NEQWM only incorporates them in a local sense (thus missing a true representation of the complete upstream history), both these models will inherently produce sub-optimal $C_f$ predictions in the separation bubble, where the accumulation of the upstream APG effects peaks. The PDE NEQWM, on the other hand, is capable of capturing these non-local effects, courtesy of the nonequilibrium terms the model includes fully  as well as a globally-connected grid in the wall parallel direction. 


\subsubsection{Case C35: Separation bubble with sweep (3D-TBL) }\label{sec:3D_bubble}

\begin{figure}[t]
\centering
\begin{subfigure}[b]{0.44\textwidth}
    \includegraphics[trim=0 180 600 200,clip,width=\textwidth]{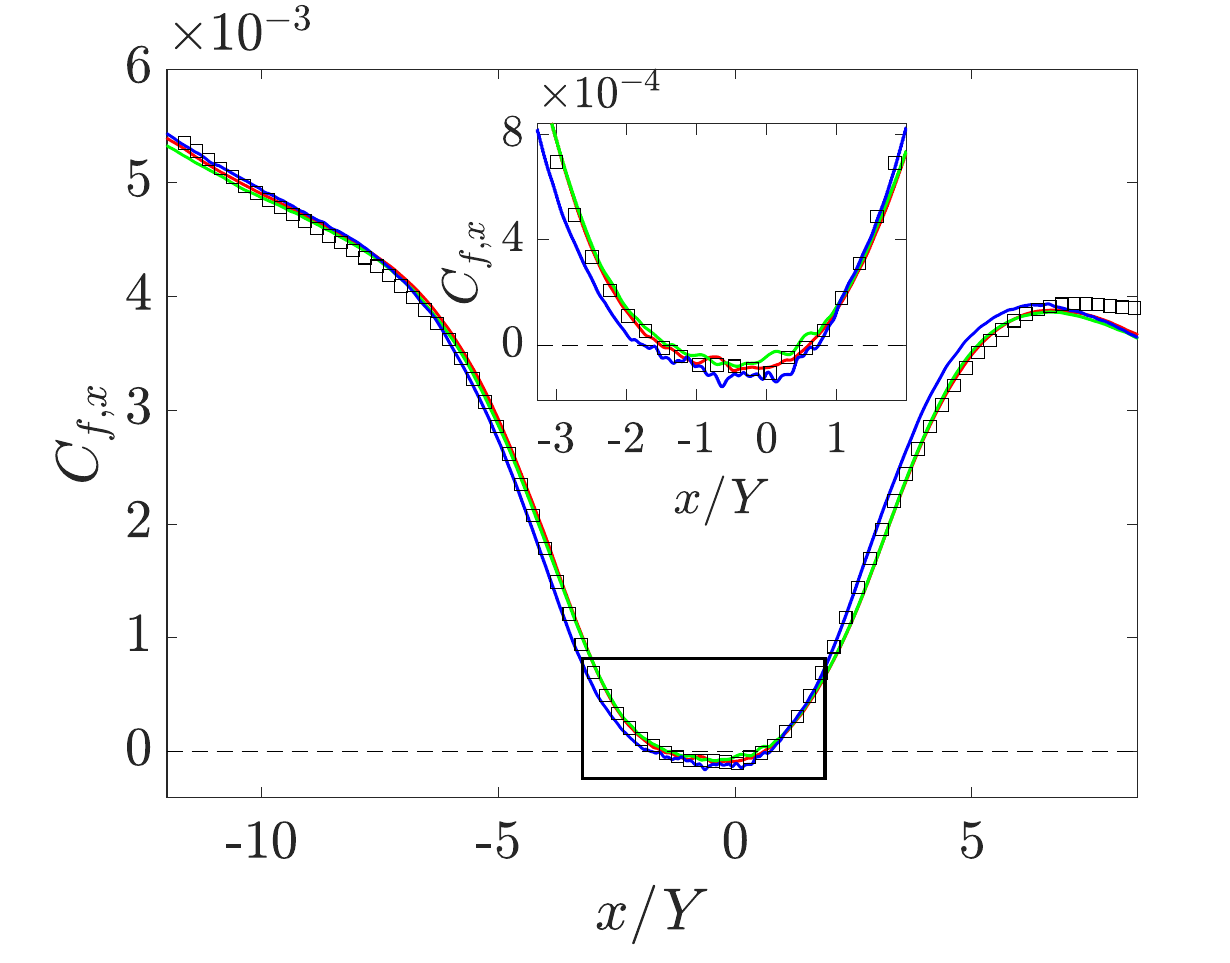}
    \caption{}
\end{subfigure}
~
\begin{subfigure}[b]{0.44\textwidth}
    \includegraphics[trim=600 180 0 200,clip,width=\textwidth]{figures/Cfx_Cfz_C35_WMLES.pdf}
    \caption{}
\end{subfigure}

\caption{Chordwise ($x$) distribution of the mean skin friction coefficient components in Case C35: $(a)$ $C_{f,x}$, $(b)$ $C_{f,z}$. Inset in $(a)$ magnifies the boxed region to visualize separation. Line colors and symbols are the same as in Fig.~\ref{fig:cf_sepbub}.} \label{fig:cf_sweep}
\end{figure}

Recall that the non-alignment of pressure-gradient and freestream vectors in Case C35 gives rise to a 3D-TBL. This affords an idealized test case to assess the predictive performance of wall models in a 3D-TBL undergoing separation over an infinite $35^{\circ}$ swept wing. In addition to the prediction of separation characteristics via $C_{f,x} = 2\tau_{w,x}/(\rho_\infty U^2_\infty)$, as in Case C0, quantities of interest (QoIs) to be predicted in Case C35 also include the direction of wall shear-stress vector (via $C_{f,z} = 2\tau_{w,z}/(\rho_\infty U^2_\infty)$), its effect on skewing of the mean-velocity profile in the inner layer, and potential modifications to the overall skewing mechanism across the entire TBL (especially in the outer layer). First, to assess the separation characteristics in Case C35, the chordwise distribution of the two skin friction components, $C_{f,x}$ and $C_{f,z}$, are shown in Fig.~\ref{fig:cf_sweep}. From Fig.~\ref{fig:cf_sweep}$(a)$, we observe two key differences in $C_{f,x}$ predictions of wall models in Case C35, compared to Case C0. First, the DNS does not exhibit a delayed recovery of $C_{f,x}$ downstream of the reattachment (observed in Case C0 between $x/Y \approx$ 3 and 6), and the wall models overall predict the correct trend in $C_{f,x}$. Second, unlike the preceding Case C0, where only the PDE NEQWM performed well in the separated region, the negative $C_{f,x}$ in the separated region is predicted reasonably by all three wall models in Case C35 (see inset in Fig.~\ref{fig:cf_sweep}$(a)$). The improved performance of EQWM and integral NEQWM is apparently attributed to the fact that the bubble height in Case C35 is almost twice as large as in Case C0 (see section~\ref{sec:RD_flow_phy_3D}). This places the matching location within the separation bubble, thus circumventing the need for a sign change in the velocity profile within the wall model, 
which plagued a major part of the separation bubble in Case C0. However, this only partially explains the observed predictions of wall models, based on their internal mechanisms. 
A plausible physics-based explanation is rooted in the downstream evolution of spanwise skin-friction component $C_{f,z}$, described in section~\ref{sec:RD_flow_phy_3D}, and the corresponding (ease of) prediction of $C_{f,z}$ for wall models. Figure~\ref{fig:cf_sweep}$(b)$ shows that, indeed, the lower-complexity models based on the log-law assumption produce equally good predictions for $C_{f,z}$ as PDE NEQWM. Recall from section~\ref{sec:RD_flow_phy_3D} that this is due to the retention (approximately) of the ZPG characteristics by the spanwise TBL in the downstream and the absence of nonequilibrium effects in the spanwise balance. \citet{Coleman2019} also pointed out that the introduction of spanwise skin-friction component in Case C35 makes the modeling of the total skin-friction magnitude less challenging for the RANS simulations. We find a similar trend for the WMLES in Case C35 because the retention of spanwise ZPG characteristics (and log law) helps the log-law-based wall models to at least predict the spanwise skin-friction component correctly.

\begin{figure}[t]
\centering
\begin{subfigure}[b]{0.44\textwidth}
    \includegraphics[trim=0 180 600 200,clip,width=\textwidth]{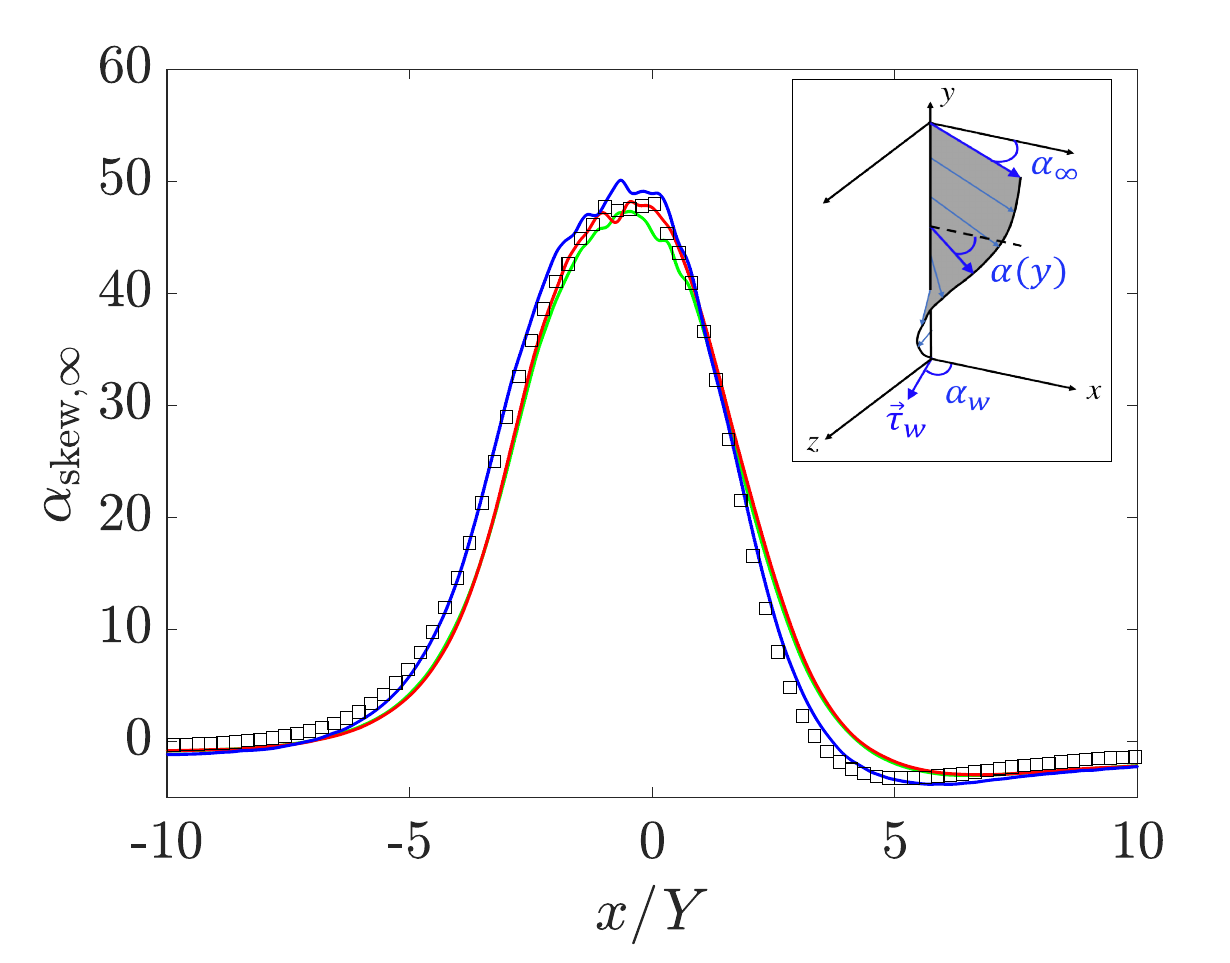}
    \caption{}
\end{subfigure}
~
\begin{subfigure}[b]{0.44\textwidth}
    \includegraphics[trim=600 180 0 200,clip,width=\textwidth]{figures/alpha_skew_C35_ystretch_and_schematic_new.pdf}
    \caption{}
\end{subfigure}

\caption{Chordwise distribution of the wall-normal skewing of horizontal mean velocity profile: ($a$) across the entire boundary-layer thickness, $\alpha_{\rm{skew},\infty} = \alpha_w - \alpha_{\infty}$, where $\alpha_{\infty} = \text{tan}^{-1}(W_e/U_e)$ is the local freestream direction; ($b$) within the wall-modeled region, $\alpha_{\rm{skew,wm}} = \alpha_w - \alpha_{h_{wm}}$, where $\alpha_{h_{wm}} =  \text{tan}^{-1}(W_{h_{wm}}/U_{h_{wm}})$ is the flow direction at the matching height. Inset in ($a$) provides a sketch of profile skewing and associated angles. Line colors and symbols are the same as in Fig.~\ref{fig:cf_sepbub}.} \label{fig:alpha_sweep}
\end{figure}

The next QoIs to be assessed are the wall-shear-stress direction and its impact on the mean three-dimensionality of flow (in the inner layer), defined as the wall-normal change in the horizontal mean-velocity direction. We quantify the effect of both by defining the skewing angle relative to the wall-shear-stress vector, $\alpha_{\rm{skew}}(y)= \alpha_w - \alpha(y)$, where $\alpha_w=\text{tan}^{-1}(\tau_{w,z}/\tau_{w,x})$ gives the wall-shear-stress direction, and $\alpha(y) = \text{tan}^{-1}(W(y)/U(y))$ gives the horizontal mean-velocity direction at a distance $y$ from the wall. Figure \ref{fig:alpha_sweep}($a$) provides a pictorial representation of these angles (in the inset),  and it also shows the chordwise distribution of $\alpha_{\rm{skew},\infty}$, which is the total profile-skewing angle across the entire boundary layer thickness. It is observed that $\alpha_{\rm{skew},\infty}$ is predicted accurately by the PDE NEQWM at all $x$, whereas, EQWM and integral NEQWM show noticeable deviation from the DNS. Note that $\alpha_{\rm{skew},\infty}$ masks the effect of wall models by including profile skewing from the outer flow, which is directly resolved on the LES grid. We expect that any departure from the DNS should be mostly concentrated in the inner layer, due to inaccuracies in the surface-flow angle prediction stemming from wall models. Indeed, Fig. \ref{fig:alpha_sweep}($b$) shows that $\alpha_{\rm{skew,wm}}$, the profile skewing within the wall-modeled region, accounts for almost all differences in $\alpha_{\rm{skew},\infty}$ between the DNS and WMLES. Figure~\ref{fig:alpha_sweep}($b$) also reveals an interesting feature of the near-wall flow in the downstream FPG: the profile skewing is negative, suggesting that $\alpha_w$ is smaller than $\alpha_{h_{wm}}$ in this region. This is explained through the delayed recovery of $C_{f,z}$ compared to $C_{f,x}$ in the FPG, as seen in Fig.~\ref{fig:cf_sweep}$(b)$.

In Fig. \ref{fig:alpha_sweep}($b$), PDE NEQWM captures the inner-profile skewing almost exactly as the DNS at all $x$, except for the downstream FPG region. This slight discrepancy is attributed to the over-prediction of $C_{f,x}$ and (especially) $C_{f,z}$ in the FPG by PDE NEQWM, as seen in Fig~\ref{fig:cf_sweep}. Note that the over-prediction of skin friction by PDE NEQWM in the FPG was described in detail in \cite{hayat2024aiaa}, where a pressure-gradient correction based on the Cebeci-Smith model \cite{cebeci1970} was applied to the wall model. In the present flow, the application of this correction, indeed, improves $\alpha_{\rm{skew,wm}}$ in the FPG region (not shown here). 
On the other hand, $\alpha_{\rm{skew,wm}}$ for EQWM is expectedly zero throughout, given its uni-directional flow assumption. Curiously, $\alpha_{\rm{skew,wm}}$ for integral NEQWM is almost zero at all $x$ (except for small non-zero values near the start of APG and the end of FPG), despite the incorporation of the spanwise momentum equation in integral NEQWM. This is explained by the fact that the viscous-sublayer formulation of integral NEQWM permits only linear (unidirectional) profiles by design, whereas the log-law part can represent some profile skewing \cite{yang15, hayat2023}. From Fig.~\ref{fig:grid_sepbub}($b$), $h^+_{wm}$ drops dramatically in the upstream and downstream vicinity of the separation bubble ($-4 \leq x/Y \leq 4$), and consequently the viscous sublayer (defined as $y^+ \leq 11$ in integral NEQWM) accounts for most of the wall-modeled region. Therefore, integral NEQWM profiles in this region are largely reduced to linear profiles. Note that this is in clear contrast to 3D-TBL in a 30$^\text{o}$ bent duct \cite{hu23}, where integral NEQWM produced improved predictions of $\alpha_{\rm{skew},\infty}$ over EQWM. This shows that by combining multiple non-equilibrium effects (3D-TBL, pressure gradient, and separation), Case C35 offers a unique test for turbulence models that highlights their vulnerabilities, which would otherwise be hidden if these non-equilibrium effects were imposed in isolation. 


\begin{figure}[t]
\centering
\begin{subfigure}[b]{0.45\textwidth}
    \includegraphics[trim=90 320 120 340,clip,width=\textwidth]{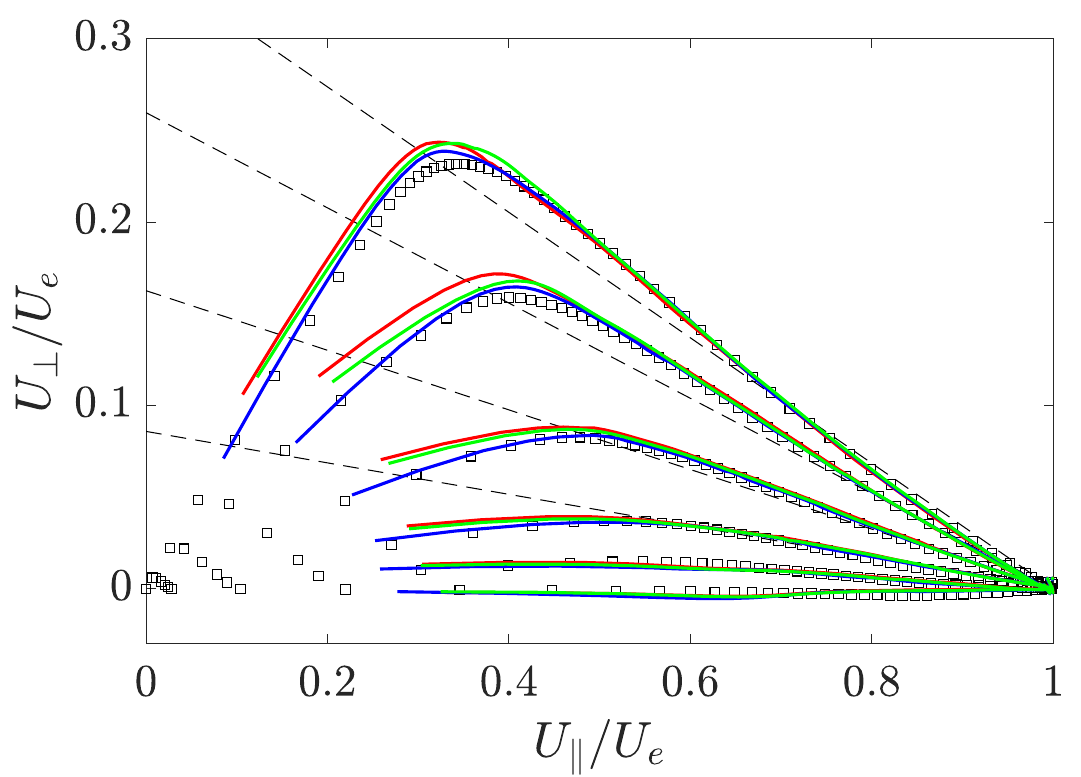}
    \caption{}
\end{subfigure}
~
\begin{subfigure}[b]{0.45\textwidth}
    \includegraphics[trim=90 320 120 340,clip,width=\textwidth]{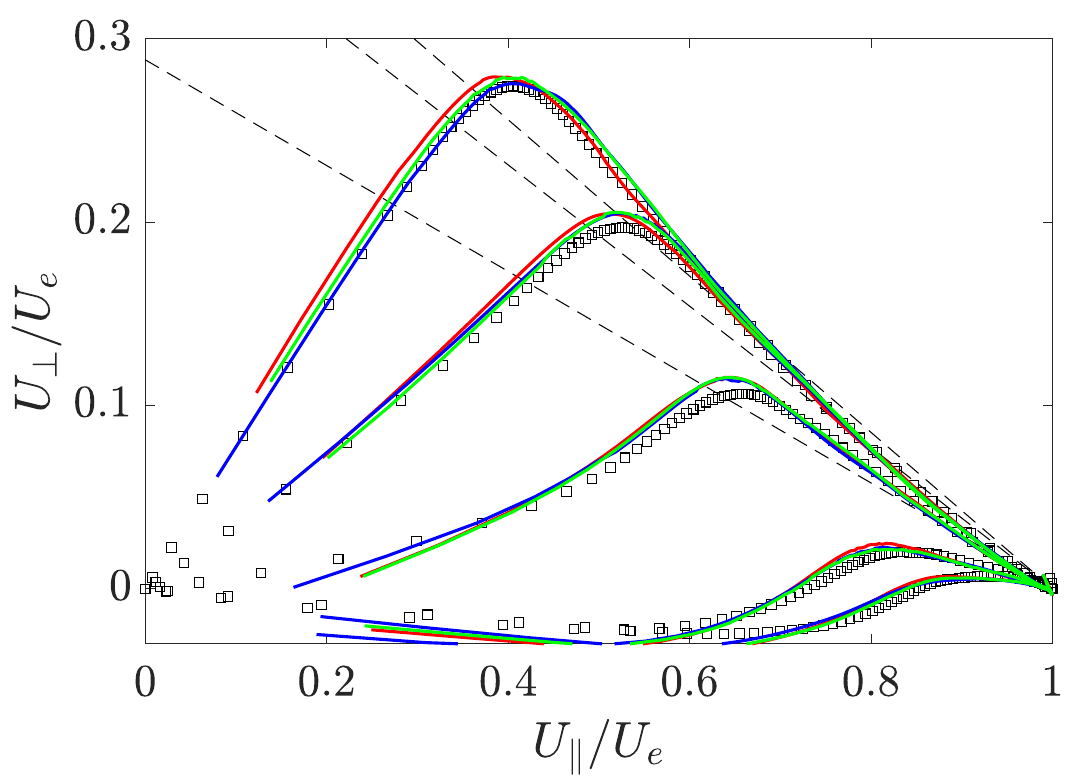}
    \caption{}
\end{subfigure}
\caption{\small  
Johnston triangular plot for Case C35, at stations ($a$) upstream of the bubble (from bottom to top: $x/Y=-8.5, -6, -5, -4, -3, -2$), and ($b$) downstream of the bubble (from top to bottom: $x/Y=1, 2, 3, 4, 7$). Dashed lines represent the SWH relation, $U_{\perp}/U_{e} = (1-U_{\parallel}/U_{e})\tan(2\theta_{SWH})$, where $U_{\parallel}$ and $U_{\perp}$ are the components parallel and perpendicular to the local edge velocity $\mathbf{U_{e}} = \overline{u}_e \mathbf{\hat{e}_{x}} + \overline{w}_e \mathbf{\hat{e}_{z}}$, $U_{e}=|\mathbf{U_{e}}|$,  $\theta_{SWH}=\sigma - \arctan(\overline{w}_e/\overline{u}_e)$ \citep{Bradshaw1987}. Symbols, DNS; solid lines, WMLES (red, EQWM; blue, PDE NEQWM; green, integral WM). Note that the left-most point in the Johnston triangular plot ($U_{\parallel}/U_{e}=0$) is at the wall and the right-most point ($U_{\parallel}/U_{e}=1$) is at the freestream.
}
\label{fig:johnston_triangle}
\end{figure}

Finally, we investigate the effects of wall-model predictions on the outer-layer profiles, or in other words, the overall skewing mechanism across the TBL. Since the 3-D flow is primarily induced by the streamwise variation of the pressure gradient (more precisely, its cross-stream component), the dominant mechanism of skewing is expected to be the inviscid skewing mechanism, where the deflection of existing mean spanwise vorticity induces non-zero streamwise vorticity \cite{Bradshaw1987}. A popular way to analyze the skewed mean profiles so induced is through the Johnston triangular plot \cite{johnston1960}, where velocity components parallel and perpendicular to the local freestream are plotted against each other. Figure~\ref{fig:johnston_triangle} shows the Johnston triangular plots at various streamwise locations, upstream and downstream of the bubble. The linear SWH lines provide a robust test for the validity of the inviscid skewing mechanism \citep{Squire1951,Hawthorne1951}. It is observed that the outer profiles (from the DNS and WMLES) follow the SWH relation far upstream and downstream of the bubble. However, as the separation bubble is approached from both upstream and downstream, the inner part of the outer profiles starts deviating appreciably from the SWH relation. This indicates that inviscid skewing is no longer the dominant skewing mechanism and that viscous (or shear)-induced skewing might be somewhat significant in the proximity of the bubble. A comparison of Johnston triangular plots between EQWM and PDE NEQWM shows that the latter predicts profiles and their skewness more accurately, especially in the $x$-vicinity of the bubble.
This indicates that the imposition of the correct wall-shear-stress direction by PDE NEQWM (as can be deduced from Fig. \ref{fig:alpha_sweep}) has a visible impact on the outer-profile prediction, further consolidating the view that shear effects are important for skewing in the vicinity of the bubble. We conjecture that the deviation from the SWH relation in the $x$-vicinity of the bubble is due to the detached (lifted) shear layer, where the significant Reynolds stresses cause stress-induced skewing in the outer layer. Therefore, the inviscid mechanism cannot solely predict the outer-profile skewing. 
One could argue that departure from SWH is due to the inaccuracy of the SWH formula for large freestream turning angles. However, the excellent agreement with SWH in previous studies featuring significantly larger turning angles than the present case, such as 3D-TBL in a 30$^\text{o}$ bent duct \cite{hu23}, makes this argument less convincing. In conclusion, the correct imposition of not only the magnitude but also the direction of wall shear stress is important for the accurate prediction of profile skewing in the separated 3D-TBL. Furthermore, contrary to the notion that the influence of Reynolds stresses on profile skewing in the outer layer is insignificant \cite{Bradshaw1987},
the present study shows this might not be true for TBLs close to separation.

\subsection{RD decomposition for wall models}\label{sec:RD_wallmodel}

In this section, we apply the RD decomposition and insights gained from analyses of the DNS datasets in section~\ref{sec:RD_flow_phy}, directly to the wall model solutions. The goal here is to identify terms in the decomposition that are  most critical to the accurate skin-friction prediction in the present flow. The RD decomposition was derived in section \ref{sec:RD_flow_phy}
without any modeling assumptions in the momentum equation, and with a limit of integration extending up to the boundary layer thickness $\delta_{99}$. Two modifications to Eq.~(\ref{eqn:RD_x_DNS}) are required to make it amenable to wall models. First, the modeled eddy viscosity term must be included in $C^*_{f2,x}$ (TKE production) and in the total shear stress $\tau$ in $C^*_{f3,x}$ on the RHS. Second, the upper limit of integration must be set to the wall-model matching height, which results in an additional term on the RHS. The resulting decomposition is then as follows (see Appendix \ref{appendix:WM_RD} for derivation),
\begin{equation}
\label{eqn:RD_x_WM}
     \begin{aligned}
        \frac{C^*_{f,x}}{2}  &=\; \overbrace{\frac{1}{U_{e}^3}\int_{0}^{ h_{wm} } \nu \left( \frac{\partial 	\langle u \rangle }{\partial y} \right)^2 dy }^{C^*_{f1,x}} \;+\; \overbrace{\frac{1}{U_{e}^3}\int_{0}^{ h_{wm} } \left(-\langle u'v' \rangle \frac{\partial \langle u \rangle }{\partial y} +  \nu_{t} \left( \frac{\partial 	\langle u \rangle }{\partial y} \right)^2  \right) dy }^{C^*_{f2,x}} \\ &+ \underbrace{\frac{1}{U_{e}^3}\int_{0}^{ h_{wm} } \left( \langle u \rangle - U_{e} \right) \frac{\partial}{\partial y}\left( \frac{\tau}{\rho} \right) dy}_{C^*_{f3,x}} +  \underbrace{\frac{1}{U_{e}^3} \left( U_{e} - U_{h_{wm}} \right) \frac{\tau_{h_{wm}}}{\rho} }_{C^*_{f4,x}}, 
    \end{aligned}
\end{equation}
\noindent where $\tau_{\text{hwm}}$ is the total shear stress at the matching height. The additional term on the RHS (hereafter referred to as $C^*_{f4,x}$) can be interpreted as the remaining kinetic energy above the wall-modeled region, out of the total kinetic energy imparted by the wall to the turbulent boundary layer. 

\begin{figure}[t]
\centering
\begin{subfigure}[b]{0.49\textwidth}
    \includegraphics[trim=110 140 780 510,clip,width=\textwidth]{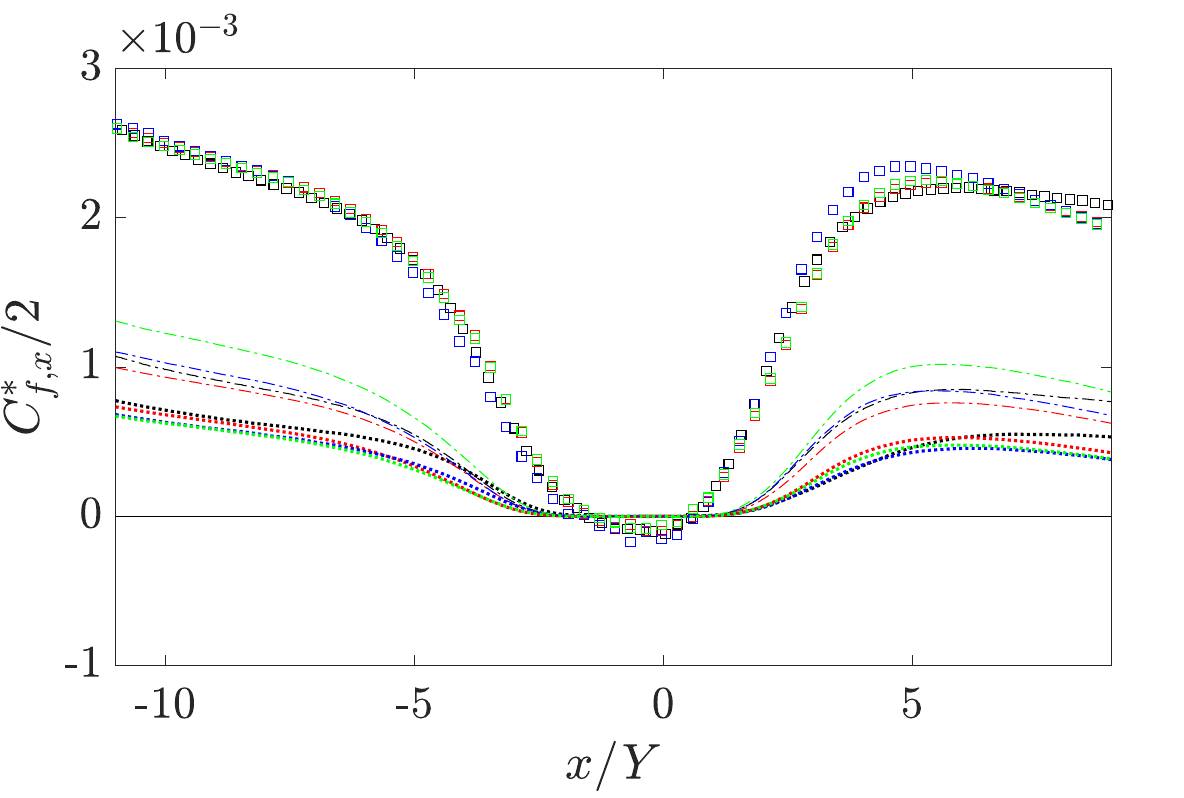}
    \caption{}
\end{subfigure}
~
\begin{subfigure}[b]{0.49\textwidth}
    \includegraphics[trim=760 140 130 510,clip,width=\textwidth]{figures/RD_Cfx_C35_WM.pdf}
    \caption{}
\end{subfigure}

\caption{Modified chordwise RD decomposition of skin friction (Eq.~(\ref{eqn:RD_x_WM})) using data from wall models and the DNS in Case C35. Symbols denote chordwise skin friction ($C_{f,x}/2$) from wall models and the DNS. Line types are used to distinguish constituent terms on the RHS of Eq.~(\ref{eqn:RD_x_WM}) (in ($a$): dashed-dotted line, $C^*_{f,1}$; dotted line, $C^*_{f,2}$; in ($b$): solid line, $C^*_{f,3}$; dashed line, $C^*_{f,4}$). Colors are used to differentiate between the DNS and different wall models (black, DNS; blue, PDE NEQWM; red, EQWM; green, integral NEQWM).} \label{fig:RD_Cfx_WM}
\end{figure}

Figure~\ref{fig:RD_Cfx_WM} shows the above decomposition for the chordwise skin friction component in Case C35. We analyze the contributing terms on the RHS in isolation. Figure~\ref{fig:RD_Cfx_WM}($a$) shows that the viscous dissipation ($C^*_{f1,x}$) and TKE production ($C^*_{f2,x}$) terms from all wall models are in reasonable agreement with the corresponding terms from the DNS, showing that the prediction of these terms does not pose a challenge to even the simple EQWM. More importantly, the differences among wall models in these terms are insignificant, indicating that $C^*_{f1,x}$ and $C^*_{f2,x}$ are not critical in differentiating the wall model predictions in the present flow. 
Note that the contribution to skin friction from Reynolds stresses ($C^*_{f2,x}$) becomes negligibly small in the vicinity of the bubble, as explained in section~\ref{sec:RD_flow_physics_2D} via the near-wall behavior of pre-multiplied $C'_{f2,x}$. Since most turbulence models (including wall models) aim to close (model) Reynolds stresses, at first we might be inclined to believe that all wall models should be able to perform well, within and close to the separation zone.
Although the comparison of overall wall-model predictions with the DNS (compare symbols in Fig.~\ref{fig:RD_Cfx_WM}) suggests this to be true, we show below that a closer inspection of individual terms on the RHS suggests otherwise. It is emphasized here that the role of a wall model in non-equilibrium TBLs is not merely to approximate near-wall turbulence. It serves the critical additional purpose of capturing nonequilibrium terms (pressure gradient and convection)
in the momentum equation, via a finer grid in the wall-normal direction and a globally (or at least locally) connected grid in the wall-parallel directions. While most wall models afford a finer wall-normal grid, only a few (such as PDE NEQWM) provide true wall-parallel connectivity. As we shall see below, the latter feature becomes crucial when considering non-equilibrium flows, especially those with upstream history effects of pressure gradient. 


Figure~\ref{fig:RD_Cfx_WM}($b$) shows that the main differences among the wall models arise in the nonequilibrium terms encapsulated by $C^*_{f3,x}$, with negative contributions in the upstream APG region and within the bubble, and a positive contribution in the FPG region. Additionally, $C^*_{f4,x}$ also shows differences among wall models, which is expected since $\tau_{h_{wm}}$ in this term is strongly linked to the gradient of the total shear stress ($\frac{\partial \tau}{\partial y}$) in $C^*_{f3,x}$ (the multiplying factor of $\tau_{h_{wm}}$ in $C^*_{f4,x}$ is almost identical among all wall models). It is observed that PDE NEQWM reproduces the spatial growth term $C^*_{f3,x}$ from the DNS quite faithfully, owing to its incorporation of non-equilibrium terms directly in its governing equations. On the other hand, the EQWM wall model with its constant-shear assumption, completely fails to represent this term. Surprisingly, integral NEQWM yields a similar result as EQWM, despite incorporating (local) non-equilibrium terms in its formulation. However, by design, these non-equilibrium effects only modify the log-law part of integral NEQWM (through a linear departure term \cite{yang15}). As explained in section~\ref{sec:3D_bubble}, the profiles from this model are reduced to linear profiles in the same regions where $C^*_{f3,x}$ from the DNS is active in Fig.~\ref{fig:RD_Cfx_WM}($b$), thereby depriving integral NEQWM of any non-equilibrium effects. Despite these discrepancies, the predicted $C_{f,x}$ distributions from EQWM and integral NEQWM agree quite well with the DNS in Fig.~\ref{fig:RD_Cfx_WM}. However, this behavior is fortuitous, 
 explained through the cancellation of errors between terms $C^*_{f3,x}$ and $C^*_{f4,x}$ in these models. That is, $C^*_{f4,x}$ is severely under/over-predicted in different regions to adjust for errors in $C^*_{f3,x}$, as seen in Fig.~\ref{fig:RD_Cfx_WM}($b$).


\section{Conclusion}\label{sec:conclusion}
Pressure-gradient-induced separation of swept and non-swept turbulent boundary layers, based on the DNS studies of Coleman \emph{et al.} (\textit{J. Fluid Mech.} (2018), vol. 847, 28-70, and \textit{J. Fluid Mech.} (2019), vol. 880, 684–706), have been analyzed for various nonequilibrium effects and their role in determining quantities of interest. In particular, this study aims to isolate physical processes critical to the predictive performance of wall models in 2D- and 3D-TBLs undergoing separation. For this purpose, a physics-based decomposition of skin friction proposed by \citet{RD2016} is employed, which is essentially a kinetic energy budget integrated across the TBL at each streamwise location in the absolute reference frame. This analysis framework called the RD decomposition, appears to be quite robust to some of the limitations of previous decomposition techniques, such as sensitivity to the boundary-layer thickness definition in the AMI framework \cite{elnahhas2022} and non-physical interpretation of critical contributing terms in the FIK identity\cite{fukagata2002}. 

Using the DNS data (Cases C0 and C35), the RD decomposition framework is first validated and then used to analyze critical flow physics. In Case C0, the streamwise evolution of 
the 
individual contributing terms 
in the RD decomposition 
shows that, while the viscous dissipation and TKE production terms dominate the balance in the upstream ZPG zone, the spatial growth and TKE production terms become the main contributors to skin friction in the APG, separated, and FPG zones. Close inspection of the wall-normal distribution of constituent terms reveals that the turbulence production (or Reynolds shear stress) from the inner layer has an insignificant contribution to skin friction in the streamwise vicinity of the separation bubble. In contrast, the spatial growth term (encapsulating nonequilibrium effects
such as pressure gradient and advection) in the inner layer appears to be the dominant contributor to skin friction within the separation bubble and in the strong APG and FPG zones immediately upstream and downstream of the bubble. For wall-stress modeling, this suggests that incorporating nonequilibrium effects in the wall model is crucial for the accurate skin-friction prediction in these zones and consequently, for the separation bubble prediction.

In Case C35, the RD decomposition correctly identifies the physical processes responsible for 
deviations of skin friction from Case C0. In particular, the enhancement of skin friction in the upstream ZPG of Case C35 is attributed to an increase in viscous dissipation and TKE production terms. Furthermore, by comparing the upstream history of Clauser pressure-gradient parameter $\beta$ between the two cases, it becomes evident that the stronger $\beta$ in Case C0 suppresses the inner layer peaks of viscous dissipation and TKE production in the upstream ZPG, resulting in the observed decrease in skin friction, which is consistent with the observations of \citet{fan2020}. In the APG region leading up to the separation bubble, the emerging outer peak of TKE production
is enhanced by the stronger $\beta$ history in Case C0. This energized outer layer (or detaching shear layer) dynamics, upstream of the separation point, appears to be the reason for the suppressed bubble size in Case C0. A stronger near-wall accumulation of upstream history effects in the spatial growth term in Case C0, is evident through an increasing imbalance of nonequilibrium effects in the log layer leading up to the bubble. The spanwise RD decomposition in Case C35 reveals that the relative contribution from each physical term remains approximately the same at all downstream locations. This indicates the retention of the upstream ZPG characteristics in the spanwise boundary layer, rendering the prediction of spanwise skin friction relatively straightforward even for simple wall models. In particular, the presence of the inner-layer turbulent energy content within the separation bubble warrants the use of wall models therein, contrasting earlier studies on 2D-TBL that downplay the importance of wall-flux modeling in the separated zones.

WMLES of Case C0 and C35 are conducted to verify various physical insights provided by the RD decomposition of the DNS. In Case C0, EQWM and integral NEQWM fail to capture the separation bubble in the wall-model domain. This is partly attributed to their internal mechanism, whereby only the separation (reversed flow) felt at the matching height manifests as separation (negative skin friction) at the wall. A more robust physics-based explanation stems from the aforementioned stronger history of $\beta$ in Case C0, causing a larger accumulation of nonequilibrium effects in the near-wall region, which the lower-fidelity wall models fail to sense. On the other hand, in Case C35, both the EQWM and integral NEQWM predict the separated region reasonably well. This is attributed to the spanwise boundary layer being turbulent within the separated region and similar to (upstream) ZPG-TBL, on which the two lower-fidelity wall models are strongly based. However, compared to PDE NEQWM, these wall models still fall short in predicting profile skewing, an important quantity of interest in 3D-TBLs. Furthermore, the Johnston triangular plot reveals that the viscous-induced skewing likely has a non-negligible effect on outer layer profile skewing, close to the separation bubble. The accurate imposition of wall-shear-stress direction by the PDE NEQWM and the corresponding accurate prediction of profile skewing away from the wall further consolidate this point.

Finally, the RD decomposition framework is directly applied to wall models, resulting in an additional term in the balance, related to the state at the matching location. While the viscous dissipation and TKE production terms are predicted reasonably well by all wall models, the spatial growth term reveals the main differences in the mechanisms of wall models. In particular, PDE NEQWM reproduces the non-constant total shear stress remarkably well in APG, separation, and FPG regions, whereas the lower-fidelity models completely fail to capture this term due to their constant-shear-stress assumption. However, a somewhat fortuitous cancellation of errors between the spatial growth term and the matching-location term in these models results in the final skin-friction prediction not being too far off from the DNS.


\section*{Funding Sources}
This investigation was funded by NASA (grant 80NSSC18M0155) and the Office of Naval Research (grant N000141712310).

\begin{acknowledgments}
The authors acknowledge valuable guidance from Gary Coleman in setting up the boundary conditions and providing additional insights on the data. We would also like to acknowledge the NASA Turbulence Modeling Resource for providing access to the DNS data on validation cases employed in this study. Computing resources supporting this work were provided by the NASA High-End Computing (HEC) Program through the NASA Advanced Supercomputing (NAS) Division at Ames Research Center. 

\end{acknowledgments}

\appendix

\section{Derivation of Eq.~(\ref{eq:VKI}) from streamwise-integrated momentum balance}\label{appendix:VKI_derivation}

Starting with the streamwise and wall-normal-integrated momentum balance used in Fig. 17($b$) of \cite{Coleman2018},

\begin{equation}
\int \frac{1}{2} C_{f} ~dx = \left(1-C_{p,\text{wall}}\right) \theta-\int(1 / 2) \delta^* ~dC_{p, \text{wall}}- \int_{0}^{Y} \frac{\langle u'u' \rangle}{U^2_\infty}~dy,
\end{equation}

\noindent where $C_{f} = \tau_w/(\rho_\infty U^2_\infty)$ is the skin friction based on $U_\infty$. Differentiating with respect to $x$,

\begin{equation}
\frac{C_{f}}{2} = \left(1-C_{p,\text{wall}}\right) \frac{d \theta}{dx} - \theta \frac{d}{dx}(C_{p,\text{wall}}) - \frac{\delta^*}{2}~ \frac{d}{dx}(C_{p,\text{wall}}) - \frac{d}{dx} \int_{0}^{Y} \frac{\langle u'u' \rangle}{U^2_\infty}~dy.
\end{equation}

\noindent Noting, from the Bernoulli's equation in the freestream, that $ 1-C_{p,\text{wall}} \approx 1-C_{p,e} = (U_e/U_\infty)^2$, the above equation reduces to,

\begin{equation}
\frac{C_{f}}{2} = \left(\frac{U_e}{U_\infty}\right)^2 \frac{d \theta}{dx} + (2+H) \theta \frac{U_e}{U^2_\infty} \frac{dU_e}{dx}- \frac{d}{dx} \int_{0}^{Y} \frac{\langle u'u' \rangle}{U^2_\infty}~dy.
\end{equation}

\noindent Rearranging the terms and using the relation $C^{*}_{f} = C_{f} (U_\infty/U_e)^2$, we get, 

\begin{equation}
 \frac{d \theta}{dx} =  \frac{C_{f}^*}{2} - (2+H) \frac{\theta}{U_e} \frac{dU_e}{dx} + \frac{1}{U^2_e} \frac{d}{dx} \int_{0}^{Y} \langle u'u' \rangle~dy,
\end{equation}

\noindent which is the same as Eq.~(\ref{eq:VKI}).

\section{Characterization of inflow, potential flow, and integrated boundary layer}\label{appendix:inlet_potential_flow}

\begin{figure}[t]
\centering
\begin{subfigure}[b]{0.48\textwidth}
    \includegraphics[trim=0 230 600 230,clip,width=\textwidth]{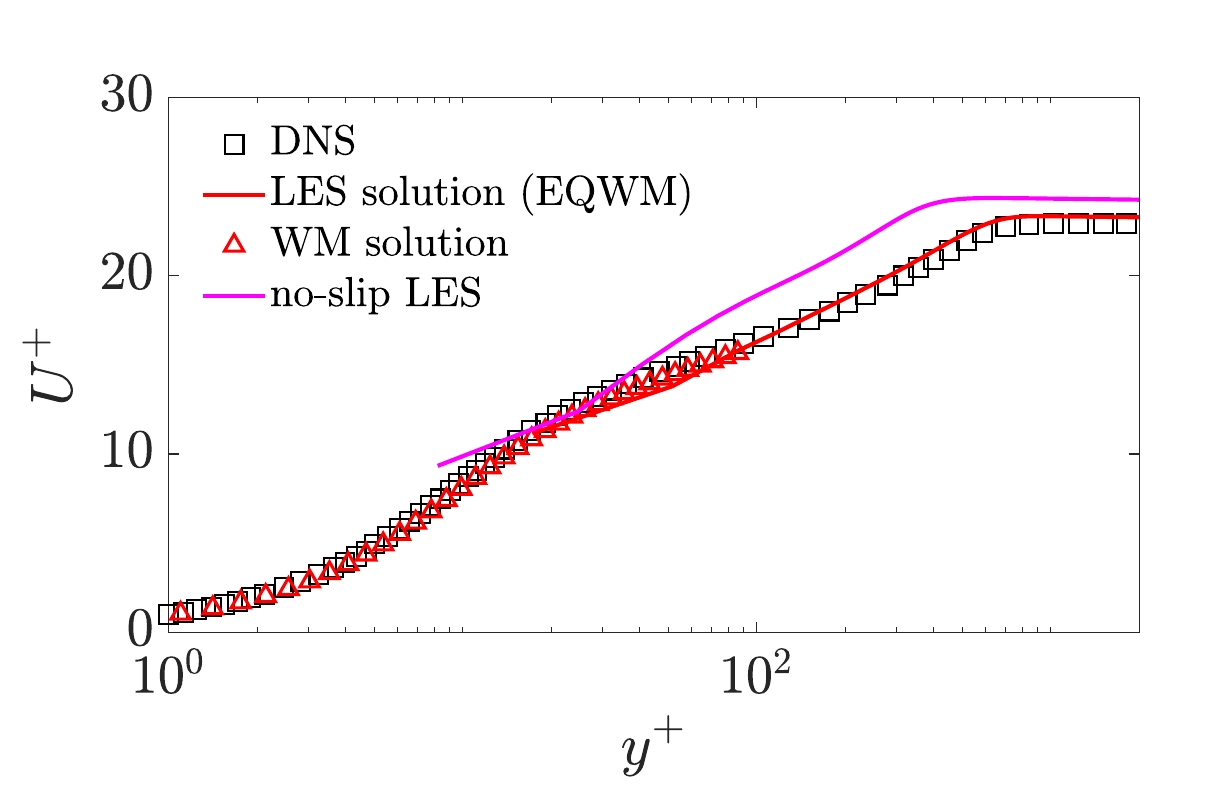}
    \caption{}
    \label{fig:Uin_C0}
\end{subfigure}
~
\begin{subfigure}[b]{0.48\textwidth}
    \includegraphics[trim=600 230 0 230,clip,width=\textwidth]{figures/Uin.pdf}
    \caption{}
    \label{fig:Cf_ZPG}
\end{subfigure}
\caption{\small  
Characterization of incoming ZPG TBL for Case C0. $(a)$ Mean streamwise velocity profile at the reference ZPG station ($x/Y = -9.5$). Squares, DNS; solid red line, fine-grid WMLES using EQWM; triangles, EQWM solution; magenta, no-slip (no-wall-model) LES. $(b)$ $C_{f}$ vs. $Re_{\theta}$ in the upstream ZPG development region. Red circles in $(b)$ are from two streamwise locations in WMLES: $x/Y = -11~\&-9.5$; squares, DNS; dashed lines, $C_{f} = 0.025 Re^{-0.25}_{\theta}$ correlation for ZPG flat-plate TBL taken from \cite{kays1980}.}
\label{fig:inflow_characterization}
\end{figure}

Characterizing the inflow TBL and matching it to the reference study is of prime importance in pressure-gradient TBLs due to the strong upstream-history dependence of these flows \citep{volino20jfm,bobke2017}. We use a synthetic turbulence generator based on a digital filter approach \citep{Klein2003} at the inflow to generate a flat-plate ZPG-TBL in the development region. The approach requires iterative guesses on the initial development length ($\ell_{x,\rm in}$), the inflow boundary-layer thickness ($\delta_{\rm 99,in}$), and the state of inflow to be prescribed at the inlet location $\ell_{x,\rm in}$ upstream of the reference ZPG location (see in Fig. \ref{fig:schematic_sepbub}).  The boundary layer statistics were reproduced reasonably well at the reference ZPG station by prescribing a flat-plate ZPG-TBL at $Re_{\theta}=900$ with $\delta_{\rm 99,in}/Y = 0.03$, and $\ell_{x,\rm in}/Y = 8.35$ and $6.35$, for C0 and C35 cases, respectively. Figure \ref{fig:inflow_characterization} shows that the mean velocity profile at the reference ZPG station in the simulation agrees well with the DNS. Furthermore, the skin friction coefficient in the development region agrees reasonably with the DNS and the characteristic $C_{f}$ distribution for flat-plate ZPG-TBL taken from a standard textbook correlation. 

 \begin{figure}[t]
\centering
\begin{subfigure}[b]{0.42\textwidth}
    \includegraphics[trim=0 180 600 200,clip,width=\textwidth]{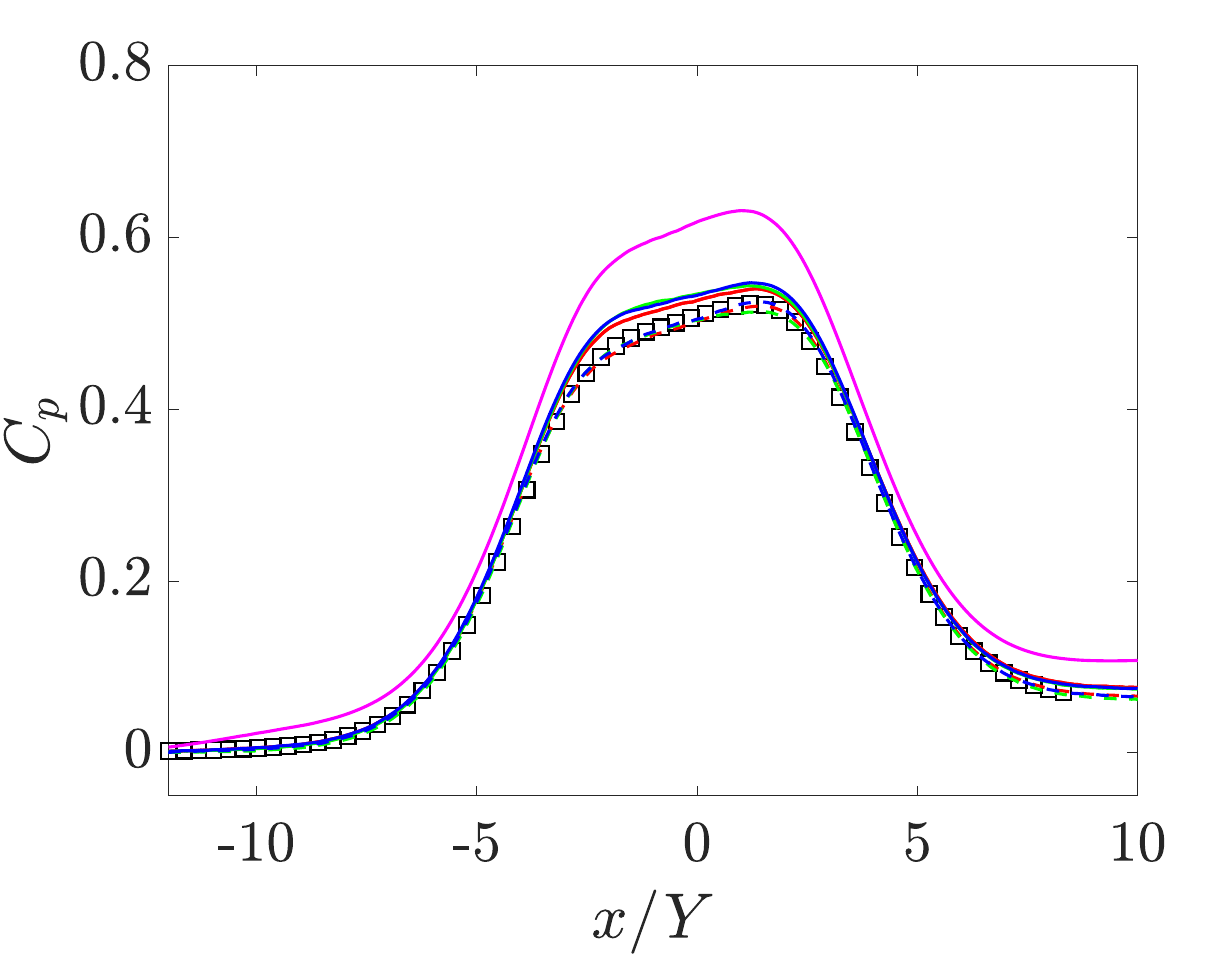}
    \caption{}
\end{subfigure}
~
\begin{subfigure}[b]{0.42\textwidth}
    \includegraphics[trim=600 180 0 200,clip,width=\textwidth]{figures/Cp_del_star.pdf}
    \caption{}
\end{subfigure}
    	
\caption{\small Potential flow characterization in Case C0 through the streamwise distribution of $(a)$ pressure coefficient, $(b)$ displacement thickness. Squares, DNS; solid lines, coarse-grid WMLES; dashed lines, fine-grid WMLES; blue, PDE NEQWM; red, ODE EQWM; green, integral NEQWM; magenta, no-slip (no-wall-model) LES.
}
\label{fig:potential_flow}
\end{figure}

The correct characterization of outer flow quantities, such as pressure coefficient and boundary-layer thickness, is critical to the accurate reproduction of flow conditions. The potential flow characterization through the streamwise distribution of pressure coefficient ($C_{p}$) is shown in Fig.~\ref{fig:potential_flow}($a$) for Case C0. The WMLES from all three wall models are in reasonable agreement with the DNS even at the coarse grid level. For the finer grid, the $C_{p}$ characterization within the bubble region improves. Furthermore, from Fig.~\ref{fig:potential_flow}($b$) the displacement thickness ($\delta^{*}$) distribution is in good agreement with the DNS at the inflow and all the downstream locations, which further consolidates the correct imposition of the potential flow and boundary conditions that drive the boundary layer. Notice that the no-slip LES fails to reproduce even the basic quantities ($C_{p}$ and $\delta^{*}$), justifying the need for wall modeling at the present Reynolds number.

\section{Wall model formulations}\label{appendix:WM_formulations}
The EQWM \citep{kawai12,bodart2012sensor} assumes a constant shear stress (equilibrium between wall shear stress and total shear stress), which is equivalent to solving only the wall-normal diffusion term in the boundary layer equations, resulting in the following system of coupled ODEs for each wall face.
\begin{eqnarray}
    \frac{d}{d\eta}\left[(\mu + \mu_{t})\frac{du_{||}}{d\eta}\right] = 0,\\
    \frac{d}{d\eta}\left[(\mu + \mu_{t})u_{||}\frac{du_{||}}{d\eta}+(\lambda + \lambda_{t})\frac{dT}{d\eta}\right] = 0,
\end{eqnarray}
\noindent where $\eta$ is the local wall-normal coordinate, $u_{||}$ is the wall-parallel velocity magnitude, $T$ is the temperature, $\mu$ is the molecular viscosity, $\lambda$ is the molecular thermal conductivity, and $\mu_t$ and $\lambda_t$ are the turbulent eddy viscosity and conductivity, respectively. The ODE is solved on a local 1-D wall-normal grid at each wall face. The wall shear stress from the wall model is assumed to be aligned with the LES velocity at the matching height. The flow is incompressible in the present study and the role of the energy equation is insignificant. The wall-model eddy viscosity $\mu_t$ is given by the following mixing-length formula,
\begin{eqnarray}
    \mu_{t} = \kappa \rho y\sqrt{\frac{\tau_w}{\rho}}D,\  D=[1-{\rm exp}(-y^+/A^+)]^2.
\end{eqnarray}

On the other hand, the PDE NEQWM \citep{park14pof,park16jcp} solves the following full 3D unsteady RANS equations on an embedded near-wall 3D grid,
\begin{eqnarray}
\frac{\partial \rho}{\partial t} + \frac{\partial \rho u_{j}}{\partial x_{j}}=0, \label{eqn:RANS1}\\
\frac{\partial \rho u_{i}}{\partial t} + \frac{\partial \rho u_{i}u_{j}}{\partial x_{j}} + \frac{\partial p}{\partial x_{i}}=\frac{\partial \tau_{ij}}{\partial x_{j}}, \label{eqn:RANS2}\\
\frac{\partial \rho E}{\partial t} + \frac{\partial(\rho E+p)u_{j}}{\partial x_{j}}=\frac{\partial \tau_{ij}u_{i}}{\partial x_{j}}-\frac{\partial q_{j}}{\partial x_{j}}, \label{eqn:RANS3}
\end{eqnarray}
\noindent where $\rho$ is the density and $u_{i}$ is the velocity component, $p$ is pressure and $E=p/[\rho(\gamma-1)]+u_{k}u_{k}/2$ is the total energy. The stress tensor and heat flux are given by, $\tau_{ij}=2(\mu+\mu_t)S_{ij}^d$ and $q_{j}=-(\lambda + \lambda_{t})\frac{\partial T}{\partial x_{j}}$.

The integral NEQWM retains the unsteady and nonequilibrium terms of the RANS equations similar to PDE NEQWM but integrates the equations in the wall-normal, and solves an ODE in time. Although the model is 3-D, the 2D formulation is summarized below to explain the working of this model. The vertically integrated momentum equation is given by,
\begin{equation}\label{eq:IWM}
\frac{\partial}{\partial t}\int_{0}^{h_{wm}} u dy + 
\frac{\partial }{\partial x} \int_{0}^{h_{wm}} u^2 dy -
U_{LES} \,\frac{\partial}{\partial x} \int_{0}^{h_{wm}} u d y = 
\frac{1}{\rho} \left[-\frac{\partial p}{\partial x} h_{wm} 
+\tau_{h_{wm}} - \tau_{w}\right],
\end{equation}
\noindent where $x$ and $y$ represent the local wall-parallel and wall-normal coordinates, $U_{LES}$ is the time-filtered velocity from the LES solution at the matching location. The integral terms in Eq.~(\ref{eq:IWM}) are evaluated by assuming an analytical composite profile for the velocity within the wall model, which has the form:
\begin{eqnarray}
u = u_{\tau} \frac{y}{\delta_{\nu}} =  \frac{u_{\tau}^2}{\nu}y, & \,\,\,\, 0 \leq y \leq \delta_{i}, \\
u = u_{\tau}\left[\frac{1}{\kappa} \log \frac{y}{h_{wm}}+C\right]+u_{\tau} A \frac{y}{h_{wm}}, & \,\,\,\, \delta_{i}<y \leq h_{wm},\label{eq:linear_departure}
\end{eqnarray}
where the unknown parameters $A$, $C$, $u_{\tau}$ and $\delta_{i}$ are determined from the solution of Eq.~(\ref{eq:IWM}) along with suitable matching and boundary conditions \cite{yang15}. This approach attempts to model the effects of nonequilibrium terms (pressure gradient and advection) through the linear departure from the log law in Eq.~(\ref{eq:linear_departure}).

\section{Derivation of RD decomposition for wall model}\label{appendix:WM_RD}
We start with the short route to RD decomposition presented in Appendix B of \cite{RD2016}. 

\begin{equation}\label{eq:appD1}
    \begin{aligned}
    \int_0^{h_{wm}}\left(\langle u\rangle-U_{e}\right) \frac{\partial}{\partial y}\left(\frac{\tau}{\rho}\right) \mathrm{d} y &= \left[\left(\langle u\rangle-U_{e}\right) \frac{\tau}{\rho}\right]_0^{h_{wm}}-\int_0^{h_{wm}} \frac{\tau}{\rho} \frac{\partial\langle u\rangle}{\partial y} \mathrm{~d} y.\\
    &= \left(U_{h_{wm}}-U_{e}\right) \frac{\tau_{h_{wm}}}{\rho} + U_{e} \frac{\tau_{w}}{\rho} - \int_0^{h_{wm}} \frac{\tau}{\rho} \frac{\partial\langle u\rangle}{\partial y} \mathrm{~d} y.
    \end{aligned}
\end{equation}

\noindent The total shear stress includes the wall-model eddy viscosity $\mu_{t}$ and is given by,

\begin{equation}\label{eq:appD2}
    \tau = \left( \mu + \mu_{t} \right) \frac{\partial\langle u\rangle}{\partial y} - \rho \langle u'v' \rangle.
\end{equation}

\noindent Substituting Eq.~(\ref{eq:appD2}) into RHS of Eq.~(\ref{eq:appD1}) and also the definition of local skin friction $C^*_{f} = 2\tau_w/(\rho U^2_e)$,

\begin{equation}\label{eq:appD3}
    \int_0^{h_{wm}}\left(\langle u\rangle-U_{e}\right) \frac{\partial}{\partial y}\left(\frac{\tau}{\rho}\right) \mathrm{d} y = \left(U_{h_{wm}}-U_{e}\right) \frac{\tau_{h_{wm}}}{\rho} + \frac{U^3_{e}}{2} C^*_{f} - \int_0^{h_{wm}}  \left[\left( \nu + \nu_{t} \right) \frac{\partial\langle u\rangle}{\partial y} - \langle u'v' \rangle \right]\frac{\partial\langle u\rangle}{\partial y} \mathrm{~d} y.
\end{equation}

\noindent Dividing by $U^3_{e}$, isolating $C^*_f/2$ on the LHS, and rearranging the rest of the terms on the RHS, we get Eq.~(\ref{eqn:RD_x_WM}):

\begin{equation}\label{eq:appD4}
    \begin{aligned}
    \frac{C^*_{f}}{2} =& \frac{1}{U^3_{e}} \int_0^{h_{wm}} \nu  \left( \frac{\partial\langle u\rangle}{\partial y} \right)^2 \mathrm{~d} y + \frac{1}{U^3_{e}} \int_0^{h_{wm}} \left[ \nu_t \left( \frac{\partial\langle u\rangle}{\partial y} \right)^2 - \langle u'v' \rangle \frac{\partial\langle u\rangle}{\partial y} \right]\mathrm{~d}y \\ +& \frac{1}{U^3_{e}}\int_0^{h_{wm}}\left(\langle u\rangle-U_{e}\right) \frac{\partial}{\partial y}\left(\frac{\tau}{\rho}\right) \mathrm{d} y + \frac{1}{U^3_{e}}  \left(U_{e} -U_{h_{wm}}\right) \frac{\tau_{h_{wm}}}{\rho}.
    \end{aligned}
\end{equation}

\bibliography{mybibfile}

\end{document}